\documentclass[twocolumn,showpacs,preprintnumbers,amsmath,amssymb,floatfix]{revtex4}

\usepackage{graphicx}
\usepackage{dcolumn}
\usepackage{bm}
\usepackage{color}
\usepackage{url}

\newcommand{\be}{\begin{equation}}
\newcommand{\ee}{\end{equation}}
\newcommand{\ba}{\begin{eqnarray}}
\newcommand{\ea}{\end{eqnarray}}
\newcommand{\Mc}{{\cal M}}
\newcommand{\Ms}{M_{\odot}}
\newcommand{\Msun}{M_{\odot}}
\newcommand{\m}{\langle}
\newcommand{\M}{\rangle}

\newcommand{\pvec}{\vec{\theta}}
\def\H{\mathcal{H}}

\def\ltsima{$\; \buildrel < \over \sim \;$}
\def\simlt{\lower.5ex\hbox{\ltsima}}
\def\gtsima{$\; \buildrel > \over \sim \;$}
\def\simgt{\lower.5ex\hbox{\gtsima}}
\begin{document}

\title{Bayesian coherent analysis of in-spiral gravitational wave signals with a detector network}

\author{J. Veitch$^{1,2}$, A. Vecchio$^1$} 
\affiliation{$^1$\,School of Physics and Astronomy, University of Birmingham, 
  Edgbaston, Birmingham B15 2TT, UK \\
  $^2$\,School of Physics and Astronomy, Cardiff University, Cardiff CF24 3AA, UK\\}
 
\begin{abstract}
The present operation of the ground-based network of gravitational-wave laser interferometers in ``enhanced'' configuration and the beginning of the construction of second-generation (or advanced) interferometers with planned observation runs beginning by 2015
bring the search for gravitational waves into a regime where detection is highly plausible. The development of techniques that allow us to discriminate a signal of astrophysical origin from instrumental artefacts in the interferometer data and to extract the full range of information
are therefore some of the primary goals of the current work. Here we report the details of a Bayesian approach to the problem of inference for gravitational wave observations using a network (containing an arbitrary number) of instruments, for the computation of the Bayes factor between two hypotheses and the evaluation of the marginalised posterior density functions of the unknown model parameters. 
The numerical algorithm to tackle the notoriously difficult problem of the evaluation of large multi-dimensional integrals is based on a technique known as Nested Sampling, which provides an attractive (and possibly superior) alternative to more traditional Markov-chain Monte Carlo (MCMC) methods. We discuss the details of the implementation of this algorithm and its performance against a Gaussian model of the background noise, considering the specific case of the signal produced by the in-spiral of binary systems of black holes and/or neutron stars, although the method is completely general and can be applied to other classes of sources. We {also} demonstrate the utility of this approach by introducing a new coherence test to distinguish between the presence of a coherent signal of astrophysical origin in the data of multiple instruments and the presence of incoherent accidental artefacts, and the effects on the estimation of the source parameters as a function of the number of instruments in the network.
\end{abstract}

\preprint{{LIGO-P0900117}}


\maketitle

\section{Introduction}
\label{s:intro}

Searches for gravitational waves are entering a crucial stage with the network of ground-based laser interferometers -- LIGO~\cite{BarishWeiss:1999, lsc-s5-instr}, Virgo~\cite{virgo} and GEO\, 600~\cite{geo} -- now fully operational and engaged in a new year-long data taking period~\cite{Smith:2009,eligo} at 
``enhanced'' sensitivity, which may allow the first direct detection of gravitational radiation. Construction has already begun for the upgrade of the instruments to advanced configuration (second generation interferometers) with installation at the sites that will start at the end of 2011~\cite{Smith:2009,advligo,advvirgo}. {When} science observations resume at much improved sensitivity {by 2015}, several gravitational-wave events are expected to be observed, opening a new means to explore a variety of astrophysical phenomena (see \emph{e.g.} Ref.~\cite{Cutler:2002me,Kokkotas:2008} and references therein).

Coalescing binary systems of compact objects -- black holes and neutron stars -- will be the workhorse source for gravitational wave observations. Ground-based laser interferometers will monitor the last seconds to minutes of the coalescence of these systems. The theoretical modelling of the (in-spiral) waveform is well in hand (see \emph{e.g.}~\cite{Blanchet:2006} and references therein), and the search algorithms are well understood~\cite{lsc-cbc-s5-12-to-18, lsc-cbc-s5y1, lsc-cbc-spins, lsc-cbc-s3s4, lsc-cbc-s2-bbh, lsc-cbc-primordial-bh, lsc-cbc-s2-bns, lsc-cbc-s1}. The detection rate for on-going searches and observations with second generation instruments is estimated to lie in the range $9\times 10^{-5}\,\mathrm{yr}^{-1} - 0.7\,\mathrm{yr}^{-1}$ and $0.2\,\mathrm{yr}^{-1} - 1000\,\mathrm{yr}^{-1}$, respectively, see~\cite{cbc-rates} for a review.
It is likely that in a few years time ground-based laser interferometers will allow us to extract a wealth of new information ranging from the formation and evolution of binary stars, the nature of precursors of (short) gamma-ray bursts, dynamical processes in star clusters, and could yield a new set of standard candles for precise cosmography.

As instruments are beginning to operate at a meaningful sensitivity from an astrophysical, cosmological and fundamental physics point of view, much emphasis is now being placed on the development of methods that offer the maximum discriminating power to separate {disturbances} of instrumental origin from a true astrophysical signal, and to extract the full range of information from the detected signals. Bayesian inference provides a powerful approach to both model selection (or hypothesis testing) and parameter estimation. Despite the conceptual simplicity of the Bayesian framework, there has been only limited use of these methods for ground-based gravitational-wave data analysis due to their computational burden, in this case related to the need to compute large multi-dimensional integrals. Additionally, the Gaussian likelihood functions considered so far do not address the instrumental glitches which are present in data from the current generation of gravitational wave detectors.

Here we present an efficient method to compute concurrently the full set of quantities at the heart of Bayesian inference: the Bayes factor between competing hypotheses and the posterior density functions (PDFs) on the relevant model parameters. 
The method is based on the \emph{Nested Sampling} algorithm~\cite{Skilling:AIP,Skilling:web,Sivia} to perform multi-dimensional integrals, that present the practical and computationally intensive challenge for the implementation of Bayesian methods.
We demonstrate the algorithm by considering multi-detector observations of gravitational waves generated during the in-spiral phase of the coalescence of a binary system, modelled using the restricted post$^{2.0}$-Newtonian stationary phase approximation, which is the waveform used so far for searches of non-spinning binary objects~\footnote{We note that searches for low-mass (that is with total mass $\le 35\,M_\odot$) binary systems in the data now collected by LIGO and Virgo are based on templates computed at the post$^{3.5}$-Newtonian order. All the results presented here apply directly to that case at no additional computational costs.}.
Initial results based on this method and applied to simplified gravitational waveforms were reported in~\cite{VeitchVecchio:2008a,VeitchVecchio:2008b}. An application to the study of different waveform approximants to detect and estimate the parameters of signals generated through the numerical integration of the Einstein's equations for the two body problem in the context of the Numerical INJection Analysis (NINJA) Project was reported in~\cite{CadonatiElAl:2009,AylottEtAl:2009,AVV:2009}. In this paper we:
\begin{itemize}
\item  Provide for the first time full details about the theoretical and technical issues on which the computation of the evidence is based;
\item  Discuss the errors associated with the computation of the integrals and the associated computational costs;
\item  Show how from the nested samples one can construct at negligible computational cost the {marginalised} posterior PDFs on the source parameters;
\item Demonstrate the performance of this technique 
{in detecting a binary in-spiral signal against a Gaussian model of background noise in coherent observations using a network of detectors, and}
by introducing a new {\emph{coherence test}} to distinguish between the presence of a coherent signal of astrophysical origin in the data of multiple instruments and the presence of incoherent accidental artefacts, and 
\item {Show the effects of the number of instruments in the network on the estimation of the source parameters. 
}
\end{itemize}
The method in its present implementation can be extended in a straightforward way to the full coalescence waveform of binary systems -- in fact the first example applied on the NINJA data set is described in~\cite{AylottEtAl:2009,AVV:2009} -- and the software is available as part of the LSC Analysis Library Applications (LALapps)~\cite{lal,lalapps}. Work is already on-going in this direction. Furthermore, the approach can also be extended to other gravitational-wave signals.

The paper is organised as follows: in Section~\ref{s:modelselection} we review the key concepts of Bayesian inference and the signal generated by in-spiralling gravitational wave signals; in Section~\ref{s:NSalgo} we describe our implementation of the numerical computation of multi-dimensional integrals for the computation of the marginal likelihood and marginalised posterior density functions based on {nested-sampling}  
the limited likelihood function; Section~\ref{s:NSimplementation} contains the implementation details of the algorithm, including a quantification of the errors that affect the results as a function of the choice of the main tuning parameters of the algorithm and the effects on the scaling of the computational costs. Section~\ref{s:results} contains the results from the applications of this method to several test cases, including a Bayesian coherence test that we introduce here for the first time.

Throughout the paper we use geometric units, in which $G = c = 1$.

\section{Bayesian inference for binary in-spirals}
\label{s:modelselection}

Statistical inference can be roughly divided into two problems: (i) Model selection, or hypothesis testing, between competing hypotheses through the computation of  
{the \emph{evidences} (or marginal likelihoods) of models}, and (ii) parameter estimation (of the unknown parameters on which a model depends). In the context of Bayesian inference, both aspects are simply tackled through an application of Bayes' theorem and the standard rules of probability theory.

While this approach is mathematically straightforward, its implementation is hampered by the need to explore large parameters spaces and  perform what are in general computationally costly, high-dimensional integrals.
For gravitational waves generated by binaries with negligible spins and eccentricity -- the case considered in this paper -- the number of dimensions of parameter space is 9, and for a generic binary system (described by general relativity) the total number of parameters increases up to 17. This technical aspect is one of the main factors that has limited the application of a Bayesian approach in a number of problems.

Model selection has been tackled through a number of techniques, including Reversible Jump MCMC~\cite{Green:1995} and thermodynamic integration~\cite{GelmanMeng:1998}. Parameter estimation is usually dealt with using MCMC methods~\cite{mcmc-in-practice, bayesian-data-analysis}, which may include advanced techniques such as parallel tempering~\cite{MarinariParisi:1992,GeyerThompson:1995} and delayed rejection~\cite{TierneyMira:1999,Mira:2001,GreenMira:2001,TVV:2009}  to enhance the exploration of the parameter space. For applications of these and other Bayesian methods in ground-based gravitational-wave observations, see \emph{e.g.}~\cite{AVV:2009,VeitchVecchio:2008b,DupuisWoan,Crab,Collaboration:2009rfa,UmstaetterTinto,Clark,Searle,vanderSluys:2007st,Rover:2006ni,RoeverMayerChristensen:2006,RoeverMayerChristensen:2007,Roever-et-al:2007,Slyus-et-al:2007}.

Nested Sampling~\cite{Skilling:AIP, Skilling:web, Sivia} is a powerful numerical technique to deal with multi-dimensional integrals. It differs from other Monte Carlo techniques such as Markov-Chain Monte-Carlo (MCMC) methods~\cite{mcmc-in-practice, bayesian-data-analysis} that are popular in applications of Bayesian inference, in that it is specifically designed to estimate the evidence integral itself, see Eq.~(\ref{e:Z}), with the marginalised posterior PDFs being optional by-products.

In this section we outline the key concepts of Bayesian inference and review the signal model -- in-spiral signals generated by non-spinning binary systems in circular orbits -- on which we concentrate the development and application of the algorithm. Many of the technical implementation aspects are associated to sampling effectively the likelihood function in the sky position parameters, and are therefore completely general for any application to short-lived bursts characterised by an arbitrary waveform.

\subsection{Model selection}
%\label{ss:overview}

In the formalism of Bayesian inference, the probability of a model or hypothesis $\H_i$, given a set of observational data ${\vec d}$ and prior information $I$ is given by Bayes' theorem,
\begin{equation}
P(\H_i|{\vec d},I)=\frac{P(\H_i|I)P({\vec d}|\H_i,I)}{P({\vec d}|I)}.
\end{equation}
In this expression, $P(\H_i|I)$ is the prior probability of $\H_i$, $P({\vec d}|\H_i,I)$ is the \emph{likelihood function} of the data, given that $\H_i$ is true, and 
\begin{equation}
P({\vec d}|I)=\sum_i P({\vec d}|\H_i,I)
\nonumber
\end{equation} 
is the marginal probability of the data set ${\vec d}$, which can only be calculated if there exists a complete set of independent hypotheses such that $\sum_j P(H_j|{\vec d},I)=1$.
Here we do not enumerate such a set of models, but we can still make comparisons between the models we do have by calculating the relative probabilities in the form of the posterior \emph{odds ratio} $O_{ij}$ between two of them,
\begin{equation}
O_{i,j}=\frac{P(\H_i|I)}{P(\H_j|I)}\frac{P({\vec d}|\H_i,I)}{P({\vec d}|\H_j,I)}=\frac{P(\H_i|I)}{P(\H_j|I)}B_{ij}\,;
\label{e:Oij}
\end{equation}
in the previous equation the normalisation factor $P({\vec d}|I)$ cancels out, and 
\begin{equation}
B_{i,j} \equiv \frac{P({\vec d}|\H_i,I)}{P({\vec d}|\H_j,I)}
\label{e:Bij}
\end{equation} 
is known as the \emph{Bayes Factor} or ratio of likelihoods.

The Bayes factors can be directly found for hypotheses which have no free parameters, but the gravitational wave signal we are modelling depends on a set of parameters, $\pvec\in\Theta$, described in Section \ref{ss:waveform}, where $\Theta$ is the parameter space.
In this case, the likelihood of the model $\H$ must be marginalised over all the parameters weighted by their prior probability distribution, giving the \emph{marginal likelihood} or \emph{evidence},
\begin{equation}
Z=P({\vec d}|\H,I)=\int_\Theta p(\pvec|\H,I)p({\vec d}|\H,\pvec,I)d\pvec,
\label{e:Z}
\end{equation}
where $p(\pvec|\H,I)$ is the prior probability distribution over the parameter space.

The integral~(\ref{e:Z}) cannot be computed analytically in all but the most trivial cases, and standard grid-based numerical approaches can take a prohibitively long time to complete when the model is high-dimensional and/or very large with respect to the posterior as is the case for gravitational-wave observations. By using the nested sampling algorithm, developed by Skilling \cite{Skilling:AIP}, we have been able to solve the problem of calculating this integral -- and as a consequence the desired Bayes factors \emph{and} the marginalised posterior density functions of the unknown model parameters $\vec{\theta}$ -- in a time that makes Bayesian techniques applicable in actual gravitational-wave search pipelines. Section \ref{s:NSalgo} provides an overview of the algorithm, and the implementation strategy that we have adopted. Further implementation details, as well as the characterisation of its accuracy in the evaluation of the evidence integral as a function of CPU time are discussed in Section~\ref{s:NSimplementation}.

\subsection{Parameter estimation}
%\label{ss:overview}

In general the hypotheses depend on a set of unknown parameters $\pvec\in\Theta$. As part of the inference process, one wants also to compute the \emph{posterior density function} (PDF)
\begin{equation}
p(\pvec|{\vec d},\H,I) = \frac{p(\pvec|\H,I) p({\vec d}|\pvec,\H,I)}{p({\vec d}|\H,I)}
\label{e:posteriorPDF}
\end{equation}
of the parameters, in this specific case of binary systems quantities such as the masses, position in the sky and distance.

The marginalised PDF on a subset $\vec{\theta}_A$ of the parameters -- our notation is $\pvec \equiv \{\pvec_A,\pvec_B\}$,  $\pvec_{A,B}\in\Theta_{A,B}$ -- is defined as
\begin{equation}
p(\pvec_A|{\vec d},\H,I) = \int_{\Theta_{B}} p(\pvec|{\vec d},\H,I) d\pvec_B\,.
\label{e:margposteriorPDF}
\end{equation}
From $p(\pvec_A|{\vec d},\H,I)$ it is then straightforward to compute \emph{e.g.} the posterior mean
\begin{equation}
\langle \pvec_A \rangle = \int_{\Theta_{A}}  \pvec_A\, p(\pvec_A|{\vec d},\H,I) d\pvec_A\,. 
\end{equation}
We will show in Section~\ref{ss:Posterior} that the Nested Sampling algorithm provides a way of computing the marginalised posterior PDFs, Eq.~(\ref{e:posteriorPDF}), with totally negligible additional computational costs from the results of the numerical evaluation of the evidence, Eq.~(\ref{e:Z}). In this respect, Nested Sampling may provide advantages with respect to more traditional MCMC algorithms.

\subsection{Target waveform}
\label{ss:waveform}

In this paper we consider observations of gravitational waves from a network (of an arbitrary number) of interferometers. The datum (in the frequency domain) at frequency $f$ from each detector that we label with $D$ is:
\begin{equation}
\tilde d^{(D)}(f) = \tilde h^{(D)}(f) + \tilde n^{(D)}(f)\,,
\label{e:d}
\end{equation}
where $\tilde h^{(D)}(f)$ and $\tilde n^{(D)}(f)$ are the gravitational wave signal and noise contribution, respectively. In Section \ref{s:results} we will consider specific choices of the interferometer network, and we will label with $D = H, L, V$ the LIGO Hanford 4-km arm instrument, the LIGO Livingston interferometer and Virgo, respectively. In practice, one works with discrete data, and we will refer to $d^{(D)}_k$ (and analogously $\tilde h^{(D)}_k$ and $\tilde n^{(D)}_k$ for the signal and noise, respectively) as the data point at discrete frequency $f_k$ of the instrument $D$. As short hand notation, we will use ${\vec d}$ to identify the whole data set from the relevant network of instruments, and to $d^{(D)}$ to the data set from a single detector $D$, so that
\begin{equation}
{\vec d} = \{d^{(H)}, d^{(L)}, ....\}\,.
\label{e:dvec}
\end{equation}

Let us consider a (geocentric) reference frame, and a gravitational wave source described by the two polarisation amplitudes $\tilde h_+(f)$ and $\tilde h_\times(f)$ located in the sky at $(\alpha,\delta)$, where $\alpha$ is the right ascension and $\delta$ the declination of the source. The signal as measured at the output of the detector $D$ is therefore
\begin{equation}
\tilde{h}^{(D)}(f) = \left[F_+^{(D)}\tilde{h}_+(f) + F_{\times}^{(D)}\tilde{h}_\times(f)\right] e^{-2\pi{}if\Delta{}t^{(D)}}\,,
\label{e:hD}
\end{equation}
where $F_+^{(D)}(\psi,\alpha,\delta;t_0)$ and $F_\times^{(D)}(\psi,\alpha,\delta;t_0)$ are the detector response functions to each polarisation, dependent on the polarisation angle $\psi$ (see \emph{e.g.} Appendix B of \cite{Anderson2001} for the definition conventions), and the time of observation $t_0$. These are computed using functions available in the LSC Algorithm Library~\cite{lal}. Given a source at location $(\alpha,\delta)$, $\Delta{}t^{(D)}(\alpha,\delta;t_0)$ is the difference in gravitational-wave arrival time between the geocentre and the detector $D$, computed with respect to a reference time $t_0$ that 
identifies the observation, see the text after Eq.~(\ref{e:psi}) below for our specific choice of $t_0$. $\Delta{}t^{(D)}$ depends on the time of observation, as for a fixed position in the sky, the signal impinges on the instruments with different relative time delays due to the Earth's rotation. By using the transformation~(\ref{e:hD}), the waveform phase which is the most expensive part of the model, needs only to be calculated once, and is then transformed to the observed signal in each detector.

In this paper, we concentrate on the in-spiral signal generated during the coalescence of a binary system of compact objects (black holes or neutron stars) of masses $m_1$ and $m_2$. Other mass parameters that we will use are the total mass $M = m_1 + m_2$, the symmetric mass ratio $\eta = m_1 m_2/(m_1 + m_2)^2$ and the chirp mass ${\cal M} = (m_1 m_2)^{3/5}/(m_1 + m_2)^{1/5}$.  We assume circular orbits and we further restrict to compact objects that are non-spinning.  We note however, that (an earlier implementation of) the approach discussed here was already successfully applied to the case of the full coalescence waveform generated by compact binaries~\cite{AylottEtAl:2009,AVV:2009}. Moreover, most of the results presented in this paper are totally general and independent of the specific waveform model, and can be applied and/or extended to any class of signals.  

The model for the gravitational-wave signal that we consider is the frequency domain, stationary phase, post$^{2.0}$-Newtonian approximation to the waveform, and more precisely the so-called ``TaylorF2'' approximant -- for an up-to-date summary of the different TaylorF/T approximants we refer the reader to~\cite{BuonannoEtAl:2009} and references therein. The waveform is therefore described by two intrinsic parameters, the two masses or any two independent combinations of them, such as ${\cal M}$ and $\eta$. We note that the specific choice of the post-Newtonian order is irrelevant for the issues discussed in this paper, as long as the waveform model used to construct the likelihood function matches the one adopted in the ``injections''\footnote{We will indulge in the LSC jargon, and use the term \emph{injection} to denote a gravitational-wave signal added to noise.} to generate synthetic data sets to explore the algorithm. As a consequence the results presented here would be essentially identical if we had adopted the post$^{3.5}$-Newtonian order which is currently used in the analysis of the LIGO/Virgo data for the current science run (S6/VSR2).
We generate the waveform directly in the frequency domain using functions of the LSC Analysis Library (LAL)~\cite{lal}. The frequency domain gravitational-wave polarisation amplitudes are given by
\begin{eqnarray}
\tilde{h}_+(f) & = & {\cal A} (1+\cos^2\iota) f^{-7/6} e^{i\Psi(f)} \,,\\
\tilde{h}_\times(f) & = & 2 {\cal A} \cos\iota f^{-7/6} e^{i\Psi(f) - i \pi/2}\,.
\end{eqnarray}
Here, the symbol $\iota$ denotes the inclination angle, defined as the angle between the line of sight to the source from the detector and the constant direction (as the objects are assumed to be non-spinning) of the orbital angular momentum. The gravitational-wave phase $\Psi(f)$ at the post$^{2}$-Newtonian order is given by
\begin{align}
\Psi({\cal M}, \eta,t_0,\phi_0;f) & = 2\pi{}ft_0-\phi{}_0
\nonumber\\
& +\psi_N(\eta)\sum_{k=0}^4\psi_k(\eta)\left(\pi M{}f\right)^{(k-5)/3}\,,
\label{e:psi}
\end{align}
where $\psi_N$ and $\psi_k$ are the standard Newtonian and post-Newtonian coefficients, whose expressions can be found in \emph{e.g.}  Ref.~ \cite{Damour}. In our implementation, $t_0$ is taken as the GPS time at the geocentre at which the frequency of the gravitational wave passes that of the nominal innermost stable circular orbit, $f_\mathrm{ISCO}=(6^{3/2}\pi M)^{-1}$, and consequently $\phi_0$ is the phase of the signal at this time.
The amplitude of the gravitational wave ${\cal A} \propto \Mc^{5/6}/D_L$ and is computed by the LAL Stationary Phase Approximation Template~\cite{lal}.
In summary, the observed signal is therefore dependent on nine quantities, which for convenience we will write as the parameter vector
\begin{equation}
\pvec=\{\Mc,\eta,t_0,\phi_0,D_L,\alpha,\delta,\psi,\iota\}\,.
\end{equation} 

Finally we discuss the assumptions on the noise $n^{(D)}$. We will make the standard assumption that the noise is a Gaussian and stationary process with zero mean and variance described through the one-sided noise spectral density $S_n^{(D)}(f)$:
\begin{eqnarray}
\m \tilde{n}(f) ^{(D)} \M & = & 0\,,
\label{e:nmean}
\\
\m \tilde{n}^{(D)}(f)\, \tilde{n}^{{(D)}^*}(f') \M & = & \frac{1}{2} \delta(f - f') S(f)\,,
\label{e:Sn}
\end{eqnarray}
where $\m.\M$ stands for the ensamble average. Under these assumptions, the likelihood of a given noise realisation $n^{(D)} = n_0$ is simply given by the multivariate Gaussian distribution
\begin{equation}
p(n^{(D)} = n_0) \propto e^{-(n_0 | n_0)/2}\,,
\label{e:Ln0}
\end{equation}
where $(.|.)$ stands for the usual inner product~\cite{CutlerFlanagan:1994}, see Eq.~(\ref{e:inner}) of the Appendix.

We will further assume that the noise in different detectors is uncorrelated, so that we generalise Eq.~(\ref{e:Sn}) to
\begin{equation}
\m  \tilde{n}^{(D)}(f)\, \tilde{n}^{{(D')}^*}(f') \M= \frac{1}{2} \delta(f - f')\delta_{DD'} S^{(D)}(f)\,.
\label{e:Sn1}
\end{equation}
{The latter assumption is appropriate for sites that are well isolated from each other, but this may not be true for the two instruments co-located at the Hanford site. Here we only consider a simulated network with no more than one instrument at any location.}
In terms of the elements $\tilde n_k$ of the discrete Fourier series of the discretely-sampled time domain data, with sampling interval $\Delta{}t$ and \emph{segment length} $T$, this is given as $\Delta{}t^2\m|\tilde{n}_k|^2\M=\frac{T}{2}S(f_k)$. {Full details of the conventions used for discretely-sampled
data are given in Appendix \ref{a:definitions}.}

\subsection{Models}
\label{ss:models}

{The problem of assessing the confidence of detection of a signal in interferometer data is the primary motivation for the Nested Sampling technique and implementation we present here.}
Translated into the Bayesian framework, {assessing the confidence of detection means computing} the Bayes factor between two hypotheses, and therefore we must specify exactly which models we are comparing. These models are the mathematical descriptions of the data $\vec{d}$, Eqs.~(\ref{e:d}) and~(\ref{e:dvec}), which either contain a gravitational-wave signal $\vec{h}(\pvec)$ parameterised by a certain vector $\pvec$  (described in section \ref{ss:waveform}), or it does not. In addition, the data also contain a contribution from instrumental noise, described by Eqs.~(\ref{e:nmean}-(\ref{e:Sn1}). 
The two models we will use can be written as:
\begin{itemize}
\item $\H{}_N$: the noise-only model,  that corresponds to the hypothesis that there is only noise (with statistical properties described in Section~\ref{ss:waveform}) present in the data set:
\begin{equation}
{\vec d} = {\vec n}\,;
\end{equation}
note that in our application we assume that the noise spectral density is known\footnote{or could be estimated using the Welch method for a real analysis} and this model has therefore no free parameters;
\item $\H{}_S$: the signal model, that corresponds to the hypothesis that the data contain noise (as before) and a gravitational-wave described by the waveform family $h(\vec{\theta})$:
\begin{equation}
{\vec d} = {\vec n} + {\vec h}(\vec{\theta})\,.
\end{equation}
\end{itemize}
Although in reality, there is also a wide range of instrumental glitches and artefacts which alters the evidence of each model in a variety of ways, initially we focus on characterising the algorithm with simulated data. Strategies for distinguishing between a coherent signal and other artefacts are discussed in Section \ref{ss:coherencetest}. 

The computation of the marginal likelihood Eq.~(\ref{e:Z}) and Bayes factor, Eq.~(\ref{e:Bij}) requires the integration of the likelihood function, $p(\vec{d}|\pvec,\H,I)$, where $\H$ is either $\H_N$ or $\H_S$, multiplied by the prior density function of the unknown parameters for the given hypothesis. We discuss the choice of prior in Section~\ref{ss:Priors}. Here we concentrate on the expression of the likelihood function. {In the case of the hypothesis $\H_N$, the likelihood function is simply
\begin{equation}
p(d^{(D)}|\pvec,\H_N,I)\propto e^{-(d^{(D)} | d^{(D)})/2}\,,
\label{e:Lnoise}
\end{equation}
see Eq.~(\ref{e:Ln0})}. 

For the hypothesis $\H_S$, the likelihood of observing a data set $d^{(D)}$ at the output of the instrument $D$ given the presence of a gravitational wave $h^{(D)}({\vec \theta})$ characterised by the parameter vector ${\vec \theta}$ is {
\begin{equation}
p(d^{(D)}|\pvec,\H_S,I)\propto e^{-(d^{(D)} - h^{(D)} | d^{(D)} - h^{(D)} )/2}\,.
\label{e:L}
\end{equation}
}{The constant of proportionality is equal in equations \ref{e:L} and \ref{e:Lnoise}, and cancels when the ratio of these quantities is taken.}
If we have a data set comprising observations from multiple interferometers, say ${\vec d}=\{{\vec d}^{H},{\vec d}^{L},\ldots\}$, the Bayesian framework allows straightforward coherent analysis. To do this, we simply write the joint likelihood of the independent datasets in all the detectors
\begin{equation}
p({\vec d}|\pvec,\H,I)=\prod_{(D)}p({\vec d}^{(D)}|\pvec,\H,I)\,,
\label{e:jointL}
\end{equation}
where $p({\vec d}^{(D)}|\pvec,\H,I)$ is either given by Eq.~(\ref{e:Ln0}) or~(\ref{e:L}).
In Appendix~\ref{a:definitions} we provide explicit expressions for the likelihood function~(\ref{e:jointL}) in the case of discrete data used for the implementation in the software code.

\section{The nested sampling algorithm}
\label{s:NSalgo}

The nested sampling algorithm is described by Skilling~\cite{Skilling:AIP} as a reversal of the usual approach to Bayesian inference, in that it directly targets the computation of the evidence integral~(\ref{e:Z}), producing samples from the posterior PDF, Eq.~(\ref{e:posteriorPDF}) of the model parameters $\pvec$ as a by-product. Although the original formulation was designed as a tool for Bayesian inference, it is actually a general method of numerical integration which could be applied to other continuous integrals. The basic algorithm, described in \cite{Skilling:AIP}, is therefore applicable to a wide range of problems, but in its generality it leaves considerable decisions to be made on the implementation, configuration and tuning to each specific application. In this section, we will review the core algorithm, Section~\ref{ss:NSgeneral}, and the processing of the output of the algorithm to extract samples from the posterior PDF which can then be used for parameter estimation, Section~\ref{ss:Posterior}. In Section~\ref{s:NSimplementation} {we
provide} detailed information on our solution of the problem of sampling the limited prior distribution, Section~\ref{ss:LimitedPrior}, and we will also examine the theoretical accuracy achievable with the algorithm. {This result will then be compared} to the practical accuracy achieved and its trade-off with computational cost in Section~\ref{ss:Errors}.

{More recently, MultiNest~\cite{FerozHobson:2007,FerozHobsonBridges:2008}, based on the technique of Nested Sampling, has been applied } to data sets primarily in the context of cosmology~\cite{FerozMarshallHobson:2008,FerozEtAl_b:2008,BridgesEtAl:2008} and particle physics~\cite{FerozEtAl_a:2008,TrottaEtAl:2008,FerozEtAl_a:2009,AbdusSalamEtAl_a:2009,AbdusSalamEtAl_b:2009,LopezFoglianiEtAl:2009}. More recently it has been used on selected mock data sets {of} the Laser Interferometer Space Antenna to search for and estimate the parameters of massive black hole binary systems characterised by high $\sim 1000$ signal-to-noise ratio~\cite{FerozEtAl_b:2009}. Our implementation is fundamentally different from MultiNest in the way in which the structure of the likelihood function is explored, and in particular it replaces the clustering algorithm or ellipsoidal rejection schemes with (non-trivial) MCMC explorations of the prior range, that are specifically tailored to the observation of gravitational-waves with a network of ground-based instruments at moderate-to-low ($\simlt 20$) signal-to-noise ratio. Moreover, given the large amount of data and the need to maximise computational efficiency while retaining the accuracy of the evaluation of the relevant integrals, our study concentrates on quantifying the errors in the evaluation of the key quantities and relating them to the computational costs.

\subsection{Computing the evidence integral}
\label{ss:NSgeneral}

The evidence $Z=P({\vec d}|\H,I)$, given in equation \ref{e:Z} is found by integrating the product of the prior distribution with the likelihood function, {in other words, $Z$ is the expectation of the likelihood with respect to the prior}. Using the product rule we can easily see the relationship between the prior, likelihood, posterior PDFs and the evidence,
\begin{eqnarray}
p(\pvec|\H,I)\times{}p({\vec d}|\pvec,\H,I)&=&Z\times{}p(\pvec|{\vec d},\H,I) \label{e:prod} \\ 
\mathrm{Prior}\times{}\mathrm{Likelihood} &=&\mathrm{Evidence}\times{}\mathrm {Posterior}. \nonumber
\end{eqnarray}
As the prior and posterior are by definition normalised, the magnitude of the evidence is governed by the likelihood function, which provides a measure of how well the data fits the hypothesis $\H$. In order to evaluate $Z$ one must sum the product on the left side of equation \ref{e:prod} at each point $\pvec\in{}\Theta$ in parameter space. In our case the parameter space is a continuous manifold, and the likelihood function is a smooth function on the manifold $\Theta$ which we integrate.
When this integral is not solvable using analytic methods, we must approximate it by using a subset of points on $\Theta$, for instance by placing a lattice on parameter space. Once such an approximation to a finite number of points is made, the result becomes subject to an integration error which is dependent on the precise means of integration.

Instead of a regular lattice of points, consider a stochastic sampling of the prior distribution to generate a basket of $N$ samples -- in the nested sampling jargon called \emph{live points} -- which we will denote $\pvec_{i}$, with $i=1\ldots{}N$. The evidence integral (equation ~\ref{e:Z}) could then be expressed as
\begin{eqnarray}
Z & = & \int_\Theta p(\pvec|\H,I)p({\vec d}|\H,\pvec,I)d\pvec\,,
\nonumber\\
& \approx & \sum_{i=1}^N{}p({\vec d}|\pvec_i,\H,I)w_i\,,
\nonumber\\
& \approx & \sum_{i=1}^N{}L_iw_i\,,
\label{e:Zapprox}
\end{eqnarray}
where the ``weight''
\begin{equation}
w_i=p(\pvec_i|\H,I)d\pvec
\label{e:w}
\end{equation} 
is the fraction of the prior distribution represented by the $i$-th sample, and $L_i \equiv p({\vec d}|\H,\pvec_i,I)$ is its likelihood. In the presence of a signal, the evidence integral is typically dominated by a small region of the prior where the likelihood is high, concentrated in a fraction $e^{-H}$ of parameter space. $H$ is called the \emph{information} in the data, subject to the particular model and parameterisation used, and is measured in nats (using base 2 instead of base $e$ would give information measured in bits, where 1\,nat = $\log_2{}e$\,bits $\approx{}$1.44\,bits). H is defined as
\begin{equation}
H=\int{}p(\pvec|\vec{d},\H,I)\log\left(\frac{p(\pvec|\vec{d},\H,I)d\pvec}{dX}\right)d\pvec,
\end{equation}
and will be used in section \ref{ss:Errors} to quantify the accuracy\,\cite{Skilling:AIP}; {$X$ is the prior mass, and is defined in Eq.~(\ref{e:Zapprox1}) below.}
If it is not known in advance where in parameter space the posterior is concentrated, approximately $e^H$ points would be needed to avoid the possibility of missing the maximum of the likelihood function using a regular grid. In the case of a compact binary in-spiral signal as observed in a network of ground-based interferometers and using the parameterisation given in section \ref{ss:waveform}, $H\gtrsim20$\,nats.
If a regular grid of points was used (assuming a uniform prior), finding the weights associated with each point $w_i=1/N$ would be simple, but the number of samples needed, $N$ becomes prohibitively large. The key concept on which the nested sampling algorithm rests is the means to calculate the $w_i$ of stochastically sampled points. By evolving the collection of $N$ points to higher likelihood areas of parameter space, the algorithm simultaneously searches for the peaks of the distribution and accumulates the evidence integral as it progresses.

\begin{figure}
\resizebox{\columnwidth}{!}{\includegraphics[width=\columnwidth]{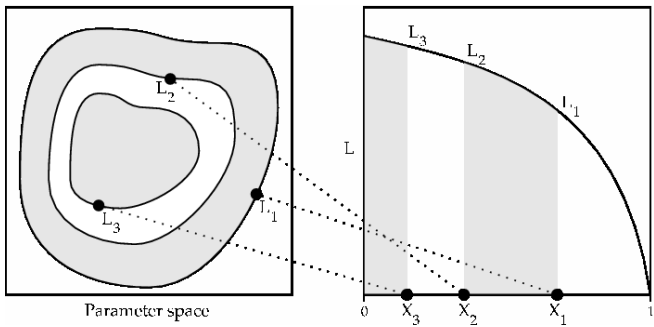}}
\caption{Each sample in the basket of live points can be thought of as lying on a contour line of equal likelihood value. Figure reproduced from \cite{Skilling:AIP}. \label{fig:contours}}
\end{figure}

In order to find the weights associated with each point $\pvec_i$, it is useful to think of each point as lying on a (not necessarily closed) contour surface of equal likelihood in the parameter space. The \emph{prior mass} -- that is the fraction of the total prior volume -- enclosed by the $i$-th contour surface is denoted $X_i$, with the lowest likelihood contour line enclosing the largest volume and the maximum likelihood point enclosing the smallest. With this definition, $X_0 = 1$. We can then think of a mapping between the contour lines in physical parameter space and the fractions of the prior $X_i$, where the likelihood $L(X)$ increases toward smaller values of $X$, as shown in figure \ref{fig:contours}{, and $\Delta{}X_i=X_{i+1}-X_i$}. The evidence, Eq.~(\ref{e:Z}) or~(\ref{e:Zapprox})  can then be expressed as the one-dimensional integral
\begin{equation}
Z=\int{}L(X)dX\approx{}\sum_i{}L(X_i)\Delta{}X_i\,.
\label{e:Zapprox1}
\end{equation}
As the inverse mapping $\pvec(X)$ is not known, the analytical integral cannot be performed. However, as we know that the prior distribution is normalised to unity, the unknown prior mass enclosed by the outermost contour through $X_1$ has a probability distribution {$P(X_1)$ which is equal to the distribution of a new variable $t_1\in{}[0,1]$, the maximum of $N$ random numbers drawn from the uniform distribution $U(0,1)$.} 
If we then replace the first point with a new point sampled from the prior distribution limited to the volume lying at higher likelihood than $L_1$, $X(L>L_1)$, we can repeat the process so that {$X_2=t_2{}X_1$ and $X_i=t_i{}X_{i-1}$, where by definition $t_i \equiv X_i/X_{i-1}$ is the shrinkage ratio. The probability of $t_i$ is $P(t_i)=Nt_i^{N-1}$, where $t_i$ is the largest of $N$ random numbers drawn from $U(0,1)$.}
The volume enclosed at each iteration therefore shrinks geometrically, ensuring the speedy convergence of the integral. The mean decrease in the volume at each iteration is 
\begin{equation}
%\langle\log{}t \rangle=-N^{-1}\,,
\mathrm{E}\left[\log{}t\right] = \int_0^1 \log{}(t)\,p(t)\,dt\,=-N^{-1}\,,
\end{equation} 
and an estimate of the statistical variance introduced by this process is
\begin{equation}
%\langle(\log{}t-\langle\log{}t\rangle)^2\rangle = N^{-1}\,.
\int_0^1\left(\log{}t-\mathrm{E}\left[\log{}t\right]\right)^2p(t)\,dt = N^{-2}\,.
\end{equation} 
The distribution of $t$ can also be sampled by generating the $N$ uniform random numbers and creating many realisations of the $t$s for each iteration in the algorithm. In our testing with this procedure, it was found that for a reasonable number of realisations, the estimated mean and variance were very close to the expected figures. Using the approximation of the mean, we can therefore write the fractional prior volumes 
\begin{equation}
\log{}X_i\approx{} - (i\pm{}\sqrt{i})/N\,,
\end{equation} 
and we use this approximation in the implementation, where we work with logarithmic quantities to overcome the huge range of the variables.

As we now have an approximation to the proportion of the prior mass remaining after the $i$-th iteration, we can assign a weight to each sample as $w_i=X_i-X_{i-1}$.

As all structure of the likelihood function in the parameter space $\Theta$ is eliminated by the mapping to $X$, the nested sampling algorithm is in principle robust against multi-modal distributions, degeneracies and problems arising from the high dimensionalality of the parameter space. However, this relies on being able to sample the likelihood-limited prior distribution effectively. This difficult problem can be solved in a variety of ways, but the approach which we have found effective is described in section \ref{ss:LimitedPrior}.

Note that if the bulk of the posterior is concentrated in a region of size $e^{-H}$ of the prior, it will take approximately $NH\pm{}\sqrt{NH}$ iterations of the geometric shrinkage to reach the zone of high likelihood. This tells us that the algorithm will take longer to compute the integral if there is a larger amount of information in the data. This means the run time of the implementation is essentially dependent on the signal to noise ratio of any found signals, as well as on the number of live points $N$ used.

\begin{figure}
\resizebox{\columnwidth}{!}{\includegraphics{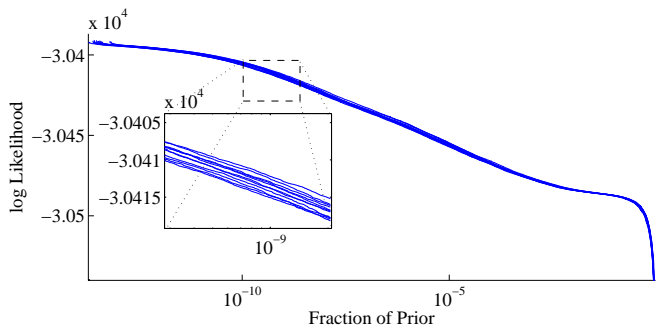}}
\caption{\label{fig:Pmap}An illustrative plot of the log likelihood against the fraction of the prior $X_i$, generated as the algorithm progresses, to be compared with figure \ref{fig:contours}. Reading this from right to left, as the fraction of the prior enclosed by the contour line decreases, the likelihood increases as the algorithm proceeds. The individual details of the distribution are smoothed out by the projection onto the $X$ parameter. The bulk of the probability from the signal occurs at around $10^{-11}=e^{-25.3}$, in good agreement with the estimated information content of $\sim{}25.7$ nats. The inset shows the slightly different results gathered by running the algorithm 10 times with a different random seed, where the precise samples used to integrate are different.}
\end{figure}

Finally, we need to specify a termination condition, upon which we decide that the integral is finished.
We could set a hard number for this, or a certain fraction of the prior, but the total number of points needed varies a great deal, particularly with the signal to noise ratio of the signal, if any.
In our implementation, we keep track of the maximum likelihood point so far discovered. The algorithm will keep running until the total evidence that would be left if all the remaining points lay at the maximum likelihood so far discovered becomes less than a certain fraction of the total evidence so far accumulated. Based upon experience, we have found that continuing while $L_{\mathrm{max}}w_i>Z_ie^{-5}$ gives consistent results.

To summarise, the algorithm can be described in pseudo-code (where $\pvec\sim{}p(\pvec|\H,I)$ means $\pvec$ is drawn from the distribution $p(\pvec|\H,I)$) as:
\begin{enumerate}
\item Draw $N$ points $\pvec_a, a\in {1}\ldots N$ from prior $p(\pvec)$, and calculate their $L_a$'s.
\item Set $Z_0=0$, $i=0$, $\log{}w_0=0$
\item While $L_\mathrm{max}w_i > Z_i e^{-5}$
\begin{enumerate}
	\item i = i +1
	\item $L_\mathrm{min}={\rm min}(\{L_a\})$
	\item $\log{}w_i=\log{}w_{i-1}-N^{-1}$
	\item $Z_i=Z_{i-1}+L_\mathrm{min}w_i$
	\item Replace $\pvec_\mathrm{min}$ with $\pvec \sim p(\pvec|\H,I) : L(\pvec)>L_\mathrm{min}$
\end{enumerate}
\item \emph{Add the remaining points:} For all $a\in {1}\ldots N$, $Z_i=Z_i+L(\pvec_a)w_i$
\end{enumerate}

With the algorithm as it is outlined above, the crucial idea is the sorting of the likelihoods so that progressively smaller contour-lines can be assigned to each of them, and the integral~(\ref{e:Zapprox}) builts up. This leads us to a natural means of \emph{parallelising the algorithm}, so that we may take advantage of multiple processors or compute nodes on a cluster. If the algorithm is run in parallel with identical data and parameters, but a different random seed (this requires no inter-node communication), the sets of samples generated will differ. If we save each sample and its likelihood value, we can then collate the results of multiple runs, and sort the resulting samples by their likelihood values. So long as the number of parallel runs $N_\mathrm{runs}$ remains constant as the integration progresses, each subsequent sample from the limited prior distribution can be treated as being part of a collection of {$N_T=\sum_{k=1}^{N_\mathrm{runs}} N_k$ samples -- where each parallel run has $N_k$ live points -- as we no longer know which sample belongs to which run.} 
This then allows us to re-apply the nested sampling algorithm as described above, but with a lower weight for each sample, substituting for $N$ the number {$N_T$.}

After applying this procedure, a more accurate estimate of the evidence integral can be obtained, using our greater number of total live points. This procedure also increases the accuracy of the evaluation of the posterior PDFs that we discuss in the next Section. It was found that this procedure allows the accuracy to scale {with the total number of parallel live points as shown} in Figure~\ref{fig:errors},
 provided that each run has a sufficiently large number of live points to avoid under-sampling of the parameter space. The issue of increasing accuracy at the expense of additional run-time is discussed further in Section \ref{ss:Errors}.

\subsection{Extracting the posterior PDF}
\label{ss:Posterior}

As the nested sampling algorithm proceeds, the list of points used in approximating the integral is stored, along with the likelihood values of each sample, the corresponding value of the parameter vector, and  $\log X_i \approx i/N$. These samples are drawn from the prior distribution, limited by a likelihood contour to a fraction $X_i$ of the full prior, meaning that the density of the samples is boosted within the contour by the probability that is excluded, as it is zero outside the contour. We can therefore write in short-hand the probability density of the $i$-th point from the nested sampling output as 
\begin{equation}
p(\pvec_i|\mathrm{NS}) = \frac{p(\pvec_i|\H,I)}{X_i}\,,
\label{e:posteriorPDF_NS}
\end{equation} 
whereas samples from the posterior PDF, Eq.~(\ref{e:margposteriorPDF}) have probability density
\begin{equation}
p(\pvec_i| {\vec d},\H,I)\propto{}p(\pvec_i|\H,I)p({\vec d}|\pvec_i,\H,I)\,.
\label{e:posteriorPDF_1}
\end{equation}
Since the nested sampling points are independent samples, they can be re-used to generate samples from the posterior PDF by re-sampling them. Substituting Eq.~(\ref{e:posteriorPDF_NS}) into Eq.~(\ref{e:posteriorPDF_1}), it is easy to see that the probabilities are related by,
\begin{equation}
p(\pvec_i| {\vec d},\H,I)\propto{}p(\pvec_i|\mathrm{NS})p({\vec d}|\pvec_i,\H,I)X_i,
\end{equation}
and so the resampling weight of each one is $\propto{}p({\vec d}|\pvec_i,\H,I)X_i$. As a consequence, the joint posterior PDF can be easily calculated by post-processing the output of the nested sampling algorithm (at negligible computational cost). Marginalised posterior PDFs, Eq.~(\ref{e:margposteriorPDF}) can then be obtained as in the case of MCMC methods, by histogramming the samples. In this way we can easily perform both evidence integrals and estimation of the posterior PDF, making both model selection and parameter estimation possible.
It is important to note that the method of extracting posterior samples using nested sampling is different to that in standard MCMC algorithms, as the algorithm is designed to move the ensemble of points uphill from a sampling of the entire prior toward the highest likelihood point, whereas MCMC can also move downhill (with probability $<1$) and requires sufficient burn-in time to fully explore the full range of the prior.
If the location of the true maximum is not known, nested sampling offers the ability to home in on the true location with its geometric shrinkage of the sampling volume, making it an excellent tool for searching the parameter space for maxima.

On the other hand, the number of posterior samples generated by the nested sampling algorithm is limited by the number of live points used, with more live points corresponding to more samples within the uppermost contour lines. In order to get the posterior sampling desired, it might be necessary to increase $N$, or run parallel computations, which has the effect of causing the algorithm to converge more slowly (but also more accurately, see Section \ref{ss:Errors}).

\section{Implementation details}
\label{s:NSimplementation}

In the previous section we have described the conceptual approach to the computation of the evidence integral using a nested sampling technique; {furthermore,} at the end of Section~\ref{ss:NSgeneral}, we have provided a pseudo-code with the key steps of the algorithm. One of the key challenges in the efficient implementation of the algorithm is to replace at each iteration 
the active point characterised by the minimum value of the likelihood function {$L_\mathrm{min}$}, drawing a sample from the prior distribution limited to the volume that satisfy the condition $L > L_\mathrm{min}$.
We do so by means of a Metropolis-Hastings Markov chain Monte Carlo with $M$ steps that proceeds as follows.
We randomly select one of the $N$ live points, corresponding to say $\pvec$, that we assume as the starting point of the Markov chain, the ``current state''. We then propose a new state ${\vec\theta}'$ drawn from a proposal distribution (also called transition kernel) $q(\pvec,\pvec')$, \emph{i.e.} the probability of $\pvec'$ given $\pvec$. The new state is accepted with probability 
\begin{equation}
\alpha_\mathrm{H}(\pvec,\pvec') =
\begin{cases}
\min\left[1,\frac{p(\pvec')q(\pvec',\pvec)}{p(\pvec)q(\pvec,\pvec')}\right] & L(\pvec')>L_\mathrm{min}\\
0 & L(\pvec')\le L_\mathrm{min}
\end{cases}
\label{e:Hratio}
\end{equation}
and equivalently the chain remains at $\pvec$ with probability $1 - \alpha_\mathrm{H}(\pvec,\pvec')$. We continue to evolve the chain to accumulate $M$ states, and the last one is set to be the new live point that replace that characterised by $L = L_\mathrm{min}$. If no points have been accepted during the $M$ proposals, then the chain will have remained at the pre-existing point, and so we must re-run the chain using a different live point as an initial state.

In Section~\ref{ss:LimitedPrior} we describe the details of the exploration of the limited prior, that is the choice of  $q(\pvec,\pvec')$; {in Section~\ref{ss:Errors}} we quantify how the errors in the evidence evaluation scale with the number of live points and MCMC elements, $N$ and $M$, respectively, and how they are related to the CPU processing time.

\subsection{Sampling the limited prior}
\label{ss:LimitedPrior}

In order to produce a new live point for each iteration of the nested sampling algorithm, it is necessary to draw a sample from the prior distribution, limited to volumes with likelihood greater than $L_\mathrm{min}$. This distribution changes from the entire prior distribution at the zeroth iteration, to a tiny fraction, typically $<10^{-10}$, of the parameter space when the posterior mode has been located, see \emph{e.g.} Figure~\ref{fig:Pmap}. In between these extremes, as $L_\mathrm{min}$ increases, it will cause ``islands'' of probability to separate from each other and disappear, as if being submerged by a rising tide. There may also be multiple maxima of similar likelihood values, and these modes are generally curved or ``banana-shaped'' in the multidimensional volume, an example of which is shown in Figure~\ref{fig:skyrings}.

Maintaining an accurate and efficient sampling of all these islands is the biggest challenge in implementing nested sampling, as it is in other Monte Carlo methods such as MCMC's. The character of these islands and their shapes will vary from problem to problem, but here we have attempted to proceed in a general way, through the use of a semi-adaptive Markov chain Monte Carlo algorithm to sample the prior, where the number of iterations of the chain $M$ can be specified. This approach was augmented with custom proposals, based on some simple intuition on the structure of the likelihood function, which were found to improve the efficiency of sampling, or alternatively the speed of chain mixing, and will be described below in section \ref{ss:MCMCcustom}.

By using an MCMC sampling of the prior, we have to choose a proposal distribution $q(\pvec,\pvec')$ which will give a decent acceptance ratio at all stages of the nested sampling. As the scale of the problem varies by ten or more orders of magnitude, this is impossible to achieve with a static choice of proposals. 
{However, we have additional information available, in the form of the location of the $N$ live points, which can help us select proposal distributions dynamically.}

As the collection of live points shrinks at each iteration, we can obtain an estimate of the size and orientation of the area we need to sample by computing the covariance matrix {$C$} of the collection of live points,
\begin{equation}
C_{ij}=\langle{}(\theta_i-\langle{}\theta_i\rangle{})(\theta_j-\langle{}\theta_j\rangle{})\rangle{}\,,
\label{e:Cij}
\end{equation}

where the indices $i,j$ denote the dimensions (9 in this specific case) of the parameter space, and $\langle . \rangle$ should be interpreted as the sample mean over the active live points. In the case of the cyclical angular parameters $\phi_0$ and $\alpha$, we need to take into account the wrapping of the boundary, and set the covariance between these and the other parameters to zero. The variances $C_{\phi_0\phi_0}$ {and $C_{\alpha\alpha}$} are then computed using the circular mean and circular difference in place of the mean and difference in the above expression~\footnote{The circular mean {of a set of angles $\theta_i$ is} defined by $ \overline{\theta} = \tan^{-1}\frac{ \overline{\sin \theta_i} } {\overline{\cos \theta_i} }  $. }.
The matrix $C_{ij}$ is re-computed at regular intervals (in practice, once every $N/10$ iterations) as the shrinking proceeds, scaled by a factor of 0.1, and used in the sampling of a multivariate student distribution, with mean centred on the previous point, and two degrees of freedom, when evolving the prior samples. 
In order to do this, a vector $\vec{v}$ is drawn from the multivariate normal distribution with mean zero and covariance matrix given by the expression (\ref{e:Cij}), then multiplied by a factor {$\sqrt{2/x}$, where $x$ is drawn from a $\chi^2$ distribution with two degrees of freedom to yield a multivariate student deviate. The vector $\vec{v}$ is then added to $\vec{\theta}$ to yield the new proposed point $\vec{\theta}' =  \vec{\theta} + \vec{v}$.}

The advantages of this somewhat ad-hoc method are that it is fast to compute, and is applicable whenever the number of live points is greater than the dimensionality of the problem to avoid a singular matrix {(which in practice is always the case). As a consequence, it} 
can be used without modification if one wants to examine a model with additional parameters, or with some of the parameters constrained. It also adapts to the shrinking volume of the limited prior distribution as the algorithm progresses. This type of proposal is used for the majority of the jumps used in the sampler. {However, there are certain types of proposals that we use specifically for short-lived bursts (such as gravitational waves generated by coalescing binaries) specifically discussed in this paper that are designed to move along the degeneracies of the distribution.}

\subsubsection{Custom proposals for sky position}
\label{ss:MCMCcustom}	

\begin{figure}
\resizebox{\columnwidth}{!}{\includegraphics{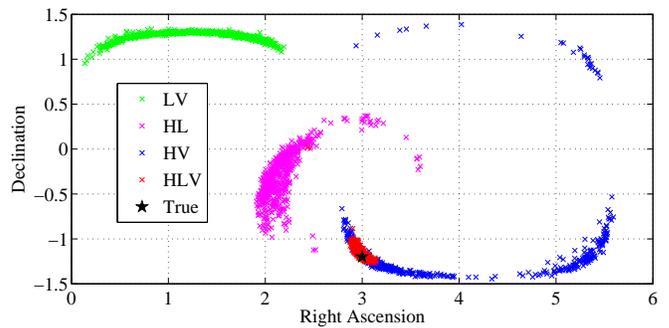}}
\caption{\label{fig:skyrings} The structure of the likelihood function can clearly be seen in the samples from the posterior PDF. These figures show the distributions obtained when using the Livingston-Virgo (\textcolor{green}{$\times$}), Hanford-Livingston (\textcolor{magenta}{$\times$}), Hanford-Virgo (\textcolor{blue}{$\times$}) and Hanford-Livingston-Virgo (\textcolor{red}{$\times$}) networks of detectors to analyse the same stretch of simulated data. The true position of the injection is marked $\star$. With two detectors, the distribution lies on the circle produced by keeping the time of arrival constant in both instruments. With a network of three detectors, the timing can be kept constant by reflecting the position of the source across the plane of the three detectors. This causes two local maxima at the two places where the three circles intersect, with one of these being the true location.}
\end{figure}

The \emph{coherent} analysis of data from interferometers at different locations on the Earth, as it is the case for the LIGO-Virgo network, offers a number of advantages that we will discuss in more detail in the next Section. Particularly relevant for the implementation details discussed here is the fact that multiple instruments allow us to reconstruct partial or full information (depending on the number of instruments in the network and their location/orientation) about the source position in the sky. Such information is encoded in the structure of the likelihood function, and its functional dependency on $(t_0,\alpha,\delta)$. However, distinctive features in the distribution also provide a challenge in exploring them in a accurate, yet efficient way. The relative timing offset between the arrival time of {a gravitational wave} 
at multiple sites encodes the majority of the information about location in the sky (of course additional information is contained in the antenna beam patterns $F_{+,\times}$). More specifically, given two detectors there exists a locus of points in the sky that traces a circle about the baseline between the two sites that yield the same time-delay at the two detectors. If one adds a third site, the three circles intersect in two points: one corresponds to the actual source position in the sky, whereas the other represents a ``mirror image'' that is located on the opposite side across the plane that passes through the three sites. The likelihood function therefore exhibits certain degeneracies or near-degeneracies in the $(t_0,\alpha,\delta)$ subspace of parameter space. This results in distributions which trace out arcs of a circle on the sky, with modulation in $t_0$, as shown in Figure \ref{fig:skyrings}.

In the exploration of the parameter space one can therefore take advantage of the known geometrical symmetries of the problem, by suitably choosing the proposals that control the geometrical parameters. It is therefore much more efficient to move ``in circles'' in the sky -- subject to the constraints that we have discussed above -- rather than to make proposals based on the distribution discussed in Section~\ref{ss:LimitedPrior}. We have indeed observed that custom-made proposals dramatically improve the performance of the algorithm, both in term of efficiency and accuracy. As the relative timing offsets between detectors provide the majority of the information about location on the sky, we therefore propose a fraction of new states of the MCMC chain by keeping the time of arrival of the signal constant in each detector. In the case of a network of two detectors, this constrains the jump to a ring, centred on the vector between the two detectors, which we sample by applying a rotation matrix with uniform random angle between 0 and $2\pi$ to the position vector of the source (and accounting for the relative rotation between the earth-fixed detector co-ordinates and the sky-fixed source co-ordinates). In the case of three detectors, in order to keep the same time of arrival in all detectors, the source must be reflected in their plane. As a consequence, if $\hat{x}$ is the current and $\hat{x'}$ the proposed Cartesian unit vector to the source, and the detectors are located at the points $\vec{x}_H$, $\vec{x}_L$, $\vec{x}_V$ in the same co-ordinate system, with a normal to their plane $\hat{n} = \left[(\vec{x}_L - \vec{x}_H) \times (\vec{x}_V - \vec{x}_H)\right]/\left|\left[(\vec{x}_L - \vec{x}_H) \times (\vec{x}_V - \vec{x}_H)\right] \right|$, 
the jump is therefore
\begin{equation}
\hat{x'}=\hat{x}-2\hat{n}\, \left|\hat{n}\cdot (\hat{x}-\hat{x}_H) \right|.
\end{equation}
In both cases, as the detectors are offset from the geocentre, and it is the time of arrival at the geocentre $t_0$ which is used as a parameter, there is also an adjustment to be made to this parameter when moving to a new sky location, $t_0'=t_0+\vec{x}_H \cdot (\hat{x}-\hat{x'})$. Here, $\vec{x}_H$ etc are measured in seconds.

The custom proposals discussed here apply to any likelihood exploration that involves short-lived (with respect to the time-scale of the Earth's rotation and orbit) bursts of gravitational waves, as it is simply connected to
the relative time delays observed between different detectors.

\subsubsection{Differential Evolution}
\label{ss:DE}

When the limited prior distribution splits into multiple isolated islands of probability, see \emph{e.g.} Figure~\ref{fig:skyrings}, the method of using the covariance matrix of the live points as a proposal distribution leads to less efficient sampling. In order to combat this effect, it is necessary to propose jumps which have a length scale characteristic of moving between, or within the multiple modes. One possible technique is to analyse the current live points as belonging to a number of clusters, then propose new states from within these clusters, which is the approach adopted by the MultiNest algorithm~\cite{FerozHobson:2007,FerozHobsonBridges:2008}.

In our implementation, we have introduced a new type of MCMC proposal which attempts to capture some of the structure of multi-modal distributions, based on a simple iteration of proposals inspired by Differential Evolution MCMC algorithms, {see \emph{e.g.} Ref.~\cite{DE}}.  From the whole set of live points $\pvec_1,\dots,\pvec_N$, we select a random point, say $\pvec_a$ that we want to evolve to point $\pvec_a'$; we then select two random existing points, say $\pvec_b$ and $\pvec_c$, such that $a\ne{}b\neq{}c$. The proposed new state is then given by 
\begin{equation}
\pvec_a'=\pvec_a+(\pvec_c-\pvec_b)\,,
\label{e:de}
\end{equation}
which is accepted with the usual Hastings ratio, Eq.~(\ref{e:Hratio}). As the probability of drawing $(b,c)$ for the random move is equal to that of drawing $(c,b)$, the move is reversible, and therefore upholds the principle of detailed balance.

When used for a fraction of the jumps (10\% in our case, chosen through trial and error to explore adjacent modes while still maintaining good diffusion of points through the standard jumps), this type of move allows proposals to be made at all the characteristic scales between the different modes in a multi-modal distribution (as well as at the scale of the width of each mode), and so increases the efficiency when such a distribution is encountered. Note that this type of proposal needs no scale to be set by the user, and is independent of the parameters used.

\subsection{Accuracy: quantifying errors}
\label{ss:Errors}

Due to the probabilistic nature of the algorithm, when computing the Bayes factor there is an associated uncertainty with the result obtained by applying the nested sampling algorithm.
As the evidence integral is written \begin{equation} Z=\sum_{i=1}^{N_\mathrm{tot}}L_iw_i,\end{equation}
there is a Poissonian uncertainty arising from the variable number of iterations needed to find the region of high posterior probability, $N_\mathrm{tot}=NH\pm\sqrt{NH}$, which gives rise to an uncertainty in $\log{}Z$ of $\pm{}\sqrt{H/N}$. In \cite{Skilling:AIP}, Skilling suggests that this error will dominate other sources of uncertainty, but makes the assumption that the sampling of the limited prior distribution is done perfectly. If the sampling is not done perfectly, for example if a small isolated mode is not properly sampled as there are no live points in its neighbourhood, an additional error will be introduced into the quantities $w_i$. Rather than attempting to derive this quantity, we have performed Monte Carlo simulations on many sets of identical data and signals, while changing the parameters of the algorithm $N$ and $M$, and examined how the distribution of estimated $Z$ values changes. By doing this, we can also explore to which value $N$ and $M$ should be set to attain a given level of accuracy in the evidence computation.
\begin{figure}
\resizebox{\columnwidth}{!}{\includegraphics{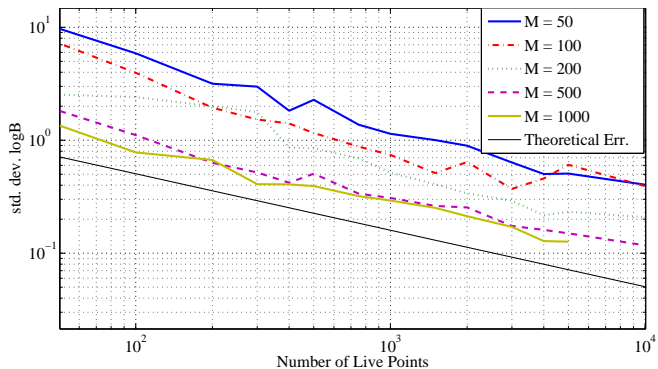}}
\caption{\label{fig:errors}The statistical error (standard deviation) {computed over 50 trials with different random seeds (identical signal and noise)} plotted against the number of live points used, using $M=50,100, 200, 500, 1000$, and the theoretical prediction of the statistical error based on Skilling's estimate. These results show that the empirical error is greater than the theoretical one, indicating an additional source of uncertainty. This is greatly reduced with the number of MCMC samples used, with chains of 1000 samples ({solid yellow line}) approaching the theoretical limit ({solid black line}). This suggests that the extra uncertainty is produced by correlation between samples, caused by less than perfect mixing in the MCMC sampling of the limited prior.}
\end{figure}
\begin{figure}
\resizebox{\columnwidth}{!}{\includegraphics{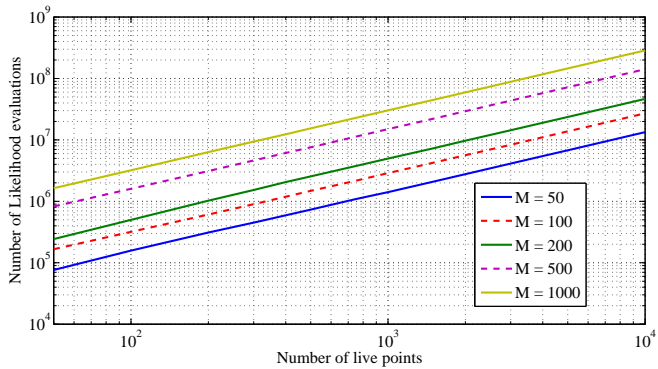}}
\caption{\label{fig:numLs}The mean number of likelihood evaluations performed in each of the trials shown in Figure \ref{fig:errors}. The number of likelihoods scales linearly with both the number of live points and the number of MCMC iterations used.}
\end{figure}

Figure \ref{fig:errors} shows the theoretical $\sqrt{H/N}$ level of uncertainty, along with the actual standard deviation of the estimates of $Z$, over 50 trials, for a range of $N$ and $M$. The actual distribution of recovered $Z$ scales as $\approx 1/\sqrt{N}$ as theoretically predicted. It is however noticeably larger than that predicted by the $\sqrt{H/N}$ error alone, and is dependent on both the number of live points and the number of MCMC samples used. We can see that when $N$ and $M$ increase, the variance decreases, suggesting that there is an additional source of error related to the sampling of the limited prior distribution. This is not entirely surprising, as the limited prior distribution consists of a number of isolated islands, between which it is difficult to move with an MCMC sampler. If there is residual correlation in the MCMC chain used to sample the parameter space, this will introduce an over-weighting of the area of over-density of samples, which, depending on whether this is a region with higher or lower likelihood will cause an increase or decrease in the $Z$ integral. As the correlation between start and end decreases with the length of the Markov chain, the added error decreases with increasing $M$. It is in fact clear that, at a given $N$, by increasing $M$ the result approaches the theoretical error.

By tuning $N$ and $M$ appropriately, we can therefore attain any desired level of accuracy, in principle, but at the expense of increasing the computational burden, as the number of likelihood evaluations is approximately proportional to the product $NM$. This can be seen clearly in Figure \ref{fig:numLs}, where we show the number of likelihood evaluations required to achieve the accuracies presented in Figure \ref{fig:errors}. The actual processing time then depends on the time taken to evaluate a single value of the likelihood function, which varies with the mass of the signal, and the length of data and sampling rate used. As an example, to compute the result with $N=500$, $M=200$, which gives an accuracy on $\log{}B$ of $\pm0.8$, $\sim3\times{}10^6$ likelihoods were evaluated. In this case the algorithm took approximately 1 hour 20 mins to complete on a 2.4 GHz Intel Xeon processor. Assuming that the overhead beyond the likelihood calculation is minimal, this gives an approximate time of 1.5ms per likelihood. The system used in these tests was a $30\,\Msun$-$1.4\,\Msun$ binary system injected into 3 data streams, but the full parameter space (see Section~\ref{ss:Priors}) was explored so that templates across the whole low-mass range were generated. It should be noted that this number is mostly dependent on the time taken to generate the waveform, which is lowest for the stationary phase approximation templates used here, but is higher for time domain templates and those requiring the numerical solution of differential equations to produce the waveform.

\section{Results}\label{s:results}

In this section we present a range of tests to demonstrate the effectiveness of a Bayesian approach in identifying gravitational-wave signals in the data from a network of interferometers and estimating the associated parameters. We first investigate the detection efficiency of our algorithm, using $\log B$ as a ``detection statistics''; we then introduce and characterise a new test to discriminate a coherent gravitational-wave at the output of multiple instruments from the presence of incoherent instrumental artefacts. We conclude by showing the impact of the number of instruments in the recovery of the source parameters.

\begin{figure}
\resizebox{\columnwidth}{!}{\includegraphics{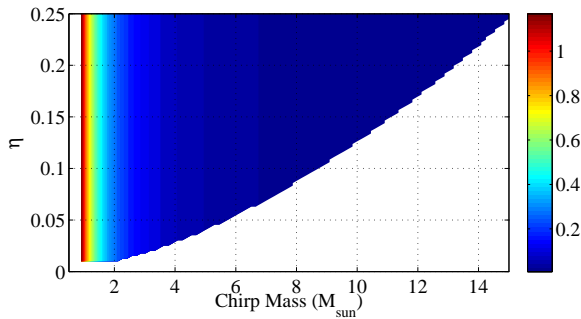}}
\caption{$p(\Mc,\eta|H_s,I)$, the prior probability distribution for the parameters $\Mc$ and $\eta$. The white area is excluded by the boundary conditions imposed on the component masses (see section \ref{ss:Priors}), and the distribution given by equation \ref{e:prM}. \label{fig:metaprior}}
\end{figure}

For the tests presented in this Section, we have chosen four systems of different combination of masses for the values used in the injections, shown in Table \ref{tab:sources}. The first three systems lie near the corners of the irregular prior, shown in Figure \ref{fig:metaprior} and discussed in Section~\ref{ss:Priors}, while the fourth lies somewhat towards the middle. Using these systems, the performance of the algorithm across different signals lying in different parts of parameter space is assessed.

\begin{table}
\begin{tabular}{|c|c|c|c|c|}
\hline
Name & $m_1$ ($\Msun$) & $m_2$ ($\Msun$) & $\Mc$ & $\eta$ \\
\hline
System 1 & 1.4 & 1.4 & 1.219 & 0.25 \\
System 2 & 30 & 1.4 & 4.727 & 0.043 \\
System 3 & 15 & 15 & 13.06 & 0.25 \\
System 4 & 25 & 5 & 9.18 & 0.139 \\
\hline
\end{tabular}
\caption{\label{tab:sources} Mass parameters of the injections used in testing.}
\end{table}

Unless stated otherwise, we run all tests using three simulated interferometers operating as a network. These are located at the sites of LIGO Hanford, LIGO Livingston and Virgo (Cascina), and have simulated noise power spectral densities chosen to correspond to the design sensitivies of each of these real instruments, as modelled by the appropriate LAL functions. The Gaussian and stationary coloured noise is then generated in the frequency domain, with the appropriate noise spectrum. A low-frequency cut-off of 50 Hz and Nyquist frequency of 1024 Hz are used throughout the analysis.

%%%%%%%%%%%%%%%%%%% PRIORS %%%%%%%%%%%%%%%

\subsection{Priors}
\label{ss:Priors}

The choice of prior distribution $p(\pvec| \H,I)$ is an important factor in Bayesian inference, and will affect the Bayes factor as it is included in the evidence integral, Eq.~(\ref{e:Z}). The prior effectively determines exactly which model is being used, by incorporating the ranges of the model parameters, and probability distribution on those parameters before the data are analysed. In the case of our implementation, the prior must be sampled using an MCMC technique, which will be more efficient when there is minimal structure in the chosen parameterisation of the signal, in the sense that fewer iterations will be required to adequately sample it.

For many of the model parameters, the choice of prior distribution is obvious:  we use an isotropic distribution for the source sky location $(\alpha, \delta)$ and the direction of the orbital angular momentum $(\iota,\psi$) over the full range on the angular parameters of the model, reflecting total ignorance and no reasons to prefer a particular geometry of a binary. We also choose a flat distribution on $\phi_0$ and $t_0$ over the range $0 \le \phi_0 \le 2\pi$ and a time interval of 100 msec, respectively. For distance and masses, we use a uniform prior on $\log{}D_L$ in the range $D_L\in{[1,100]}$\,Mpc, a uniform prior in $\eta$ and a prior on chirp mass of the form $p(\Mc|I) \propto \Mc^{-11/6}$. As we desire to test our approach on the mass region covered by the LIGO-Virgo low-mass searches for in-spiral {signals~\cite{lsc-cbc-s5-12-to-18,lsc-cbc-s5y1}}, limits were imposed directly on the component masses, such that $m_1,m_2\le{}35\Ms$, where by our convention $m_1\ge{}m_2$. We also place a lower limit on the mass ratio, such that of $\eta>0.01$, and on the chirp mass, $\Mc>0.87$, to ensure that the waveforms generated are non-zero and of a length suitable for our analysis.
These constraints result in a convoluted shape for the allowed regions of parameter space in the $(\Mc,\eta)$ plane, shown in Figure \ref{fig:metaprior}. The specific choice of the distance and mass priors is determined by the need to ensure the accuracy of the integration, which we now discuss. 

Within the core sampler, we have changed the variable used to $\log\Mc$, in order to reduce the range of the prior density, leading to better sampling of the space than using $\Mc$ itself. The prior probability density function we use on $\log\Mc$ is based on an approximation to the Jeffreys prior, $p(\pvec|\H,I)\propto\sqrt{\det \Gamma(\pvec)}$, where $\Gamma(\pvec)$ is the Fisher information matrix, {defined as
\begin{equation}
\Gamma_{ij}(\pvec)=\left(\frac{\partial_i\tilde{h}(\pvec)}{\partial\theta_i}\left|\frac{\partial_j\tilde{h}(\pvec)}{\partial\theta_j}\right.\right).
\label{e:fim}
\end{equation}
}
This type of prior is used when there is no information about the parameters of the signal at all, and should therefore be invariant under a change of co-ordinates \cite{jeffreys,Amari}. 
{For simplicity, in our initial implementation we ignore the correlations between the chirp mass and the other parameters and we just take the leading order Newtonian quadrupole approximation of the in-spiral waveform $\tilde h(f)$ to compute the scaling of the prior. Under these assumptions we obtain
\begin{eqnarray}
\frac{\partial{}\tilde{h}(f)}{\partial{}\log\Mc}& = &-\frac{5i}{4}(8\pi\Mc f)^{-5/3}\tilde{h}(f)
\end{eqnarray}
and the prior is therefore
\begin{eqnarray}
p(\log\Mc|I)&\propto&\sqrt{\left(\frac{\partial\tilde{h}(\pvec)}{\partial\log\Mc}\left|\frac{\partial\tilde{h}(\pvec)}{\partial\log\Mc}\right.\right)}\,,\nonumber \\
&\propto& \Mc^{-5/6}\,,\label{e:prlogM}\\
p(\Mc|I)&\propto&\Mc^{-11/6}\,. \label{e:prM}
\end{eqnarray}
}

The Jeffreys prior is based on the notion that one should assign equal probabilities to equal volume elements in the parameter space of the signal. Here the Fisher matrix is used as a metric, allowing us to calculate the volume element at each point. The Jeffreys prior encodes the fact that the volume element of a curved parameter space may vary with respect to the parameterisation, and so the density of templates is greater at lower chirp masses.
Failure to account for this will result in an under-sampling of certain regions of parameter space, which will increase the chances of failing to detect a signal there (or increase the number of live points needed for a given probability of detection). This effect is most significant in the $\Mc$ parameter, which is why we have found it necessary to include it here, however a full calculation of the Fisher metric in the $\Mc,\eta$ space would further improve the sampling, and is a goal for future development of this work.

Note that the use of this prior, as with the one for distance, ignores any available information about the mass distribution of neutron stars and black holes, and focusses simply on the detection of the signal with unknown parameters. 

\subsection{Detection efficiency}

\begin{figure}
\resizebox{\columnwidth}{!}{\includegraphics{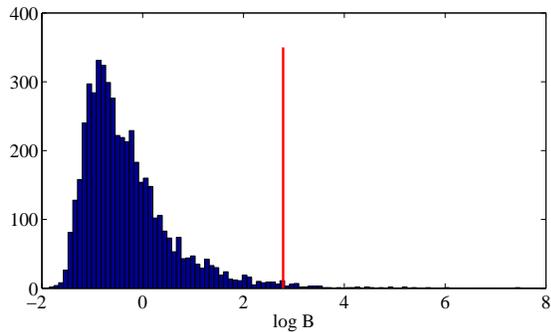}}
\caption{\label{fig:noisedist}The distribution of the log Bayes factor  for 5\,000 runs on synthetic Gaussian noise. The vertical solid line represents the 1\% false alarm rate threshold of $\log{}B_\mathrm{S,N}=2.786$.}
\end{figure}

In order to test the detection efficiency of our implementation, we have chosen to  
{treat} the Bayes factor {of the signal vs noise hypotheses
\be
B_\mathrm{S,N} = \frac{P({\vec d}|\H_S,I)}{P({\vec d}|\H_N,I)}\,,
\label{e:Bsn}
\ee
see Eq.~(\ref{e:Bij}) and Section~\ref{ss:models},} as a detection statistic. By performing a large number of runs on a signal-free dataset, we can find the distribution of $\log B_\mathrm{S,N}$ in the absence of a signal, and therefore choose a threshold value which will give a certain false alarm rate. {This is the same approach that we have adopted in Refs. \cite{VeitchVecchio:2008a}}, however {in this paper we consider coherent observations with multiple interferometers.}
Using this threshold, we can therefore decide whether or not {the analysis of a data set which contains an actual injection yields a detection or not.}
To achieve this, we analysed 5000 different realisations of Gaussian noise, and obtained the distribution of $\log B_\mathrm{S,N}$. This is shown as a histogram in Figure \ref{fig:noisedist}, where the vertical line represents the threshold $\log{}B_\mathrm{S,N}=2.786$ corresponding to a false alarm rate of 1\% for any single trial.

Taking this threshold, we then classify each result as detected if $\log{}B>2.786$, and otherwise not.
The detection efficiency is then assessed by performing 50 injections of each test signal at varying signal-to-noise ratios (by changing the distance to the source) and determining the fraction which are detected vs those which are not.
\begin{figure}
\resizebox{\columnwidth}{!}{\includegraphics{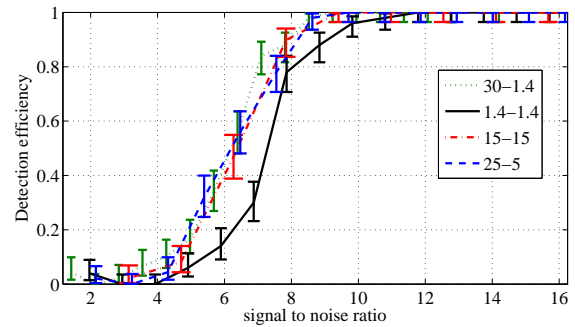}}
\caption{\label{fig:Deteff}Detection efficiency curve for the nested sampling algorithm for the four test systems as a function of the coherent {network signal-to-noise ratio}, with a detection threshold of $\log{}B_\mathrm{S.N}>2.786$, corresponding to a 1\% false alarm rate, {see Figure~\ref{fig:noisedist}}. Error bars indicate the 67\% probability interval assuming a binomial distribution for the results of the 50 trials.}
\end{figure}
This allows us to build up a detection efficiency curve for each of the test systems, given the desired false alarm rate, which is shown in Figure \ref{fig:Deteff}. From these results, we can see that the $30-1.4\,\Msun$, $15-15\,\Msun$ and $25-5\,\Msun$ systems show consistency in their chance of being detected as a function of signal-to-noise ratio, following a typical sigmoid curve with a transition zone between signal-to-noise ratios 4 and 8 where there is an intermediate chance of detection, explained by the different noise realisations. Each of these curves crosses the 50\% detection efficiency at approximately {signal-to-noise ratio of} 6.5, and approaches 100\% detection efficiency with a {signal-to-noise ratio} above 8.
In contrast, the algorithm performs slightly more poorly in the detection of the binary neutron star system with $1.4-1.4\,\Msun$ component masses, with 50\% detection efficiency at {signal-to-noise ratio} of 7.5 and a wider zone of transition. An examination of the raw Bayes factors output by the algorithm for this system indicated that the detected binary neutron star signals are allocated Bayes factors consistent with other signals of the same signal-to-noise ratio, but that there is a larger fraction of sources which are not detected, producing a Bayes factor consistent with noise. This may be due to the sampler failing to identify the correct region of parameter space, as the parameter volume of signals at low mass is considerably smaller, and therefore has a lower probability of being found by a probabilistic algorithm.

It is suggested that improved performance could be obtained for these systems by incorporating the full metric into the calculation of density required in the $\Mc,\eta$ subspace, which would then distribute the samples more appropriately. However, the results broadly show that the algorithm is capable of correctly analysing and detecting signals at an SNR comparable to existing methods. Other recent work has shown that prior distributions may be found which balance the need for efficient detection with astrophysical prior information \cite{Roever:Priors}.

\begin{figure}
\resizebox{\columnwidth}{!}{\includegraphics{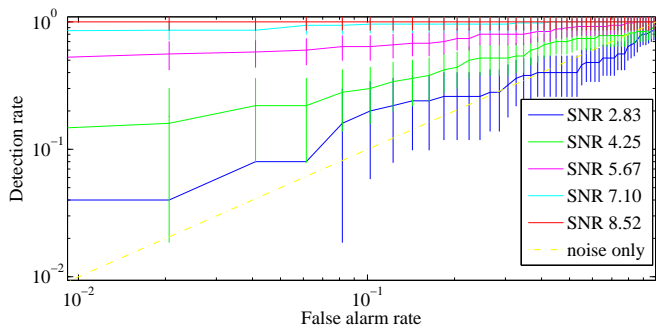}}
\caption{\label{fig:ROC}Receiver Operation Characteristics curve for the nested sampling algorithm, obtained with 50 trials at each {signal-to-noise ratio for the $30\,\Msun - 1\,\Msun$ test binary}.}
\end{figure}

Using the distribution of Bayes factors produced by the noise-only runs, we can also examine the relationship between false alarm rate and detection efficiency. For a choice of signal-to-noise ratio, the threshold of detection is varied, causing a change in the number of detected signals but also the number of false alarms, which can be plotted in a receiver operations curve {(ROC), as shown in Figure \ref{fig:ROC}}. In this figure, we have used system 1, see Table~\ref{tab:sources} with component masses $30\,\Msun$ and $1.4\,\Msun$ to produce the ROC curves for SNRs below, in and above the transition region. These results were produced with 50 independent trials at each optimal SNR, and so the error bars shown represent the Poissonian error of $50^{-1/2}$. 

\subsection{Coherent Analysis}\label{sec:coherent}

Every gravitational wave search critically relies on multiple (at least two) instruments in order to make confident detections of astrophysical signals. 
Multiple interferometers are beneficial in two main ways.
Firstly, the signal in each detector can be cross-checked against those observed in the others, giving us an additional way of isolating a real gravitational wave which must yield consistent observations in all the instruments; in fact, the searches for short-lived gravitational waves  employ one or more ``detection confidence tests'', see \emph{e.g.} \cite{Gouaty:2008} and references therein. Secondly, if the interferometers are not co-located and aligned (as is the case of the ground-based network currently in operation) then they will see a different projection of the strain tensor of the passing gravitational wave; in turn, this allows a better estimation of the parameters of the incoming signal, and can break degeneracies between parameters present with only one data set.
{In particular, simultaneous observations with three or more instruments }
allows us to determine the geometry of the source, such as its distance and location in the sky.

Using the mathematical and computational frame-work developed and described above, we can address both these points in a natural way. In this section we propose and demonstrate a new \emph{coherence test} which can discriminate between coherent and incoherent signals, making optimal use of the {data} and we demonstrate the improvement in parameter estimation when using a detector network.

\subsubsection{Coherence Test}\label{ss:coherencetest}

Having a network of detectors is essential to discriminate a real astrophysical signal from the background of spurious noise events which resemble gravitational wave signals, i.e. those with $B_\mathrm{S,N}\gg{}1$ in the language of model selection. We know that a real gravitational wave signal must be observed in all the detectors with compatible estimates for the physical parameters,
and relative time delays that are consistent with the location of the instruments on Earth and a point on the celestial sphere.
On the other hand, instrumental glitches will appear independently in each detector. To be more specific, the observed data in each detector must be consistent with the physical gravitational wave: the different characteristics of each detector, including their instantaneous noise levels and orientations mean that the signal-to-noise ratios will vary between them.

There will naturally be times when glitches occur simultaneously in multiple detectors. When this happens, we can use the network in a coherent manner to test whether the event is consistent with a coherent gravitational wave, or more likely to be a gravitational-wave-like glitch occurring independently in each detector. Translated into our framework of inference, we want to compare the two following hypotheses:
\begin{itemize}
\item Coherent model, $\H_\mathrm{coh}$: The datasets ${\vec d} = \{d^{(1)}, d^{(2)}, \ldots{}, d^{(N_D)}\}$ from each of the $N_D$ detectors contain a coherent gravitational-wave signal described by the same polarisation amplitudes $\tilde h_+(f; \pvec)$ and $\tilde h_\times(f; \pvec)$ with the same parameters $\pvec$. The posterior probability of the coherent hypothesis is as before
\begin{equation}
P(\H_\mathrm{coh}|{\vec d}) = \frac{P(\H_\mathrm{coh})}{P({\vec d})} Z_\mathrm{coh}\,,
\label{e:Pcoh}
\end{equation}
where the evidence is
\begin{equation}
Z_\mathrm{coh} = \int_\Theta{}p(\pvec|\H_\mathrm{coh}) p({\vec d}|\pvec,\H_\mathrm{coh})d\pvec
\end{equation}
and the joint likelihood of the observation ${\vec d}$ is {$p({\vec d}|\pvec,\H_\mathrm{coh})=\prod_i^{N_D} p(d^{(i)}|\pvec,\H_\mathrm{coh})$, see Eq.~(\ref{e:jointL}).}
\item Incoherent model, $\H_\mathrm{inc}$: The data set at each instrument, $d^{(1)}$, $d^{(2)}$, $\ldots{}, d^{(N_D)}$ contains independent gravitational-wave-like glitches, characterised by the parameters $\pvec^{(1)}$, $\pvec^{(2)}$, \ldots, $\pvec^{(N_D)}$, in general different for each detector. In this case, assuming that the data and signals at each instrument are independent, the marginal likelihood of the model factorises into the marginal likelihoods of each signal in the relevant detector, and the posterior probability is
\begin{eqnarray}
P(\H_\mathrm{inc}|{\vec d}) & = & \frac{P(\H_\mathrm{inc})}{P({\vec d})} Z_\mathrm{inc} \nonumber\\
& = & \frac{P(\H_\mathrm{inc})}{P({\vec d})} \prod_{i = 1}^{N_D} Z^{(i)}
\label{e:Pincoh}
\end{eqnarray}
where the evidence for the signal in each detector is
\begin{equation}
Z^{(i)} = \int_{\Theta^{(i)}} p(\pvec^{(i)}|\H_\mathrm{inc})\,p(d^{(i)} | \pvec^{(i)},\H_\mathrm{inc})d\pvec^{(i)}\,;
\end{equation}
in the equation above $p(d^{(i)} | \pvec^{(i)},\H_\mathrm{inc})$ is the likelihood of the data set $d^{(i)}$ at the $i-$th instrument output, characterised by the parameter vector $\pvec^{(i)}$ defined over the space $\Theta^{(i)}$.
\end{itemize}
In order to distinguish between these possibilities, we need to compute the odds ratio, Eq.~(\ref{e:Oij}), between the coherent and incoherent model
\begin{equation}
O_\mathrm{coh,inc} = \frac{P(\H_\mathrm{coh})}{P(\H_\mathrm{inc})}B_\mathrm{coh,inc}
\end{equation}
where the Bayes factor is
\begin{equation}
B_\mathrm{coh,inc}=\frac{Z_\mathrm{coh}}{Z_\mathrm{inc}} = \frac{Z_\mathrm{coh}}{\prod_i^N Z^{(i)}}\,,
\label{eqn:Bcoh}
\end{equation}
see Eqs~(\ref{e:Pcoh}) and~(\ref{e:Pincoh}). Essentially the test computes the difference between the integral of the product and the product of the integrals for each dataset. The incoherent model used here might be regarded as the worst possible type of glitch, in that it models a disturbance which appears exactly as a real gravitational wave would in a single detector. As usual the evaluation of the odds ratio requires to specify the prior odds, which is a subjective matter, and here we simply concentrate on the Bayes factor.

Why does this work? If we think of the parameter space of the coherent model as being embedded in the larger parameter space $\{\pvec^{(1)},\pvec^{(2)}, \ldots, \pvec^{(N_D)}\}$ of the incoherent model, we see that it lies in the nine dimensional subspace where $\pvec^{(1)} = \pvec^{(2)} = \ldots = \pvec^{(N_D)}$. If a coherent signal is present, the distribution in the incoherent space will be peaked on or near this subspace, which will intersect a relatively large total probability mass. As the coherent subspace has fewer dimensions, in this case $9^{N_D - 1}$, its prior is smaller and the model is more predictive. Loosely stated, this means that it will gain whenever its prediction is correct over the more general incoherent model. This is the so-called Occam factor which arises from comparing models with different predictive power.

The larger space of the incoherent model will also capture a greater amount of evidence from the data when a large glitch is present which causes a high likelihood in many parts of the parameter space, as any signal which is not completely parameterised by the chosen parameterisation will produce a broader peak on that manifold (such a glitch may provide a high evidence value when compared with the noise model only, but contains less information about the parameters). In this case (unless there is by chance a coherent glitch in the other detectors) the same argument will apply and cause the test to discriminate against the coherent signal model.

Performing model selection with the Bayes factor, {Eq.~(\ref{eqn:Bcoh})}, will then give us the optimal means of distinguishing a coherent gravitational wave from incoherent glitches which have a significant component which looks like an in-spiral signal. This can be thought of as an enhanced coincidence check, which uses all the parameters to check for consistency in the observed signals, and also incorporates a check for a better explanation as simultaneous glitches.

\begin{figure}
\resizebox{\columnwidth}{!}{\includegraphics{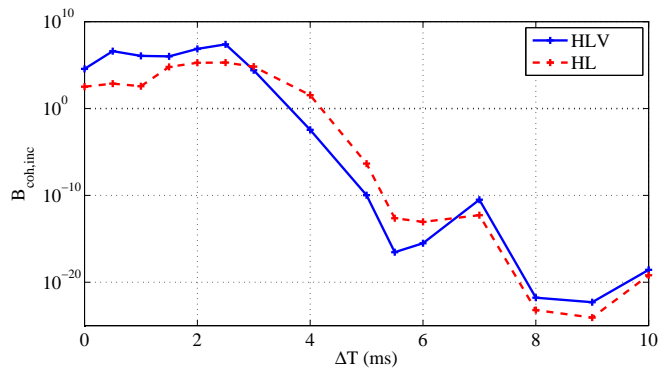}}
\caption{\label{fig:coh} Bayes factor {$B_\mathrm{coh,inc}$} between the coherent and incoherent model, {Eq.~(\ref{eqn:Bcoh})} with varying {relative} time delay between {the arrival time of a gravitational wave at different} interferometers. In solid blue, the value of $B_\mathrm{coh,inc}$ when the network of {the LIGO-Hanford, LIGO-Livingston and Virgo instruments (HLV)} was used, and in dashed red the network of {the two LIGO interferometers (HL)} . The time delay in milliseconds is applied equally between Hanford and Livingston, and Livingston and Virgo. The signal used had component masses $3\Msun$ and $4\Msun$, and a {coherent} network signal to noise ratio of 17.8. The curve shows a strong fall in coherence probability above {a time-delay of $\approx 3.5$ msec}, allowing us to rule out a single coherent signal.}
\end{figure}

Here we provide two examples to test this technique. Firstly we consider how the test performs in the analysis of data sets that contain gravitational-wave signals that have identical parameters, but are characterised by an unphysical time-offset; in the second case we analyse the case in which a gravitational wave is present only in the data of one of the instruments of the network. 

For the first test, we inject a signal characterised by identical physical parameters -- we choose $M_1 = 4\,M_\odot$, $M_2 = 3\,M_\odot$ and random position, orientation and distance such that the network optimal SNR is 17.8 in the case of a coherent injection -- into the {simulated} data streams of LIGO-Hanford, LIGO-Livingston and Virgo; we first perform the injection coherently in the data of the three interferometers, and then we repeat it by introducing a non-physical time shift $\Delta{}T \le 10$ ms in the time of coalescence at the different sites (the coherent injection case corresponds therefore to $\Delta{}T = 0$). The noise realisation is identical in each case; however, due to the slightly offset time of arrival of the signals and therefore the slightly different sum of data and signal, there is some spread in the recovered evidences. We compute $B_\mathrm{coh,inc}$, Eq.~(\ref{eqn:Bcoh}) as a function of $\Delta{}T$, considering the case in which the analysis is carried out using the two LIGO instruments (HL-network) and the three-instrument (HLV-network). The results are shown in Figure~\ref{fig:coh}, where we plot the Bayes factor against the time shift $\Delta{}T$. As expected, when the signal is injected with $\Delta{}T=0$ the evidence favours the coherent model strongly, by a factor of $\sim{}36\,461$ in the case of the HLV-network and $\sim{}328$ in the HL-network.
As the time shift $\Delta{}T$ increases {(the light-travel-time between Hanford and Livigston is $\approx 10$ msec)}, the evidence switches to favouring the incoherent model at around 4 msec in both HLV and HL networks, and rapidly decreases to strongly exclude the coherent model with a Bayes factor of $<{}10^{-10}$.
With a large separation between the signals, the incoherent model becomes very strongly preferred, by a factor up to $\sim{}2\times{}10^{22}$ in the case of the HLV and $\sim{}1\times{}10^{24}$ in the case of the HL network. {One would naively assume that HLV would yield more stringent rejection than HL, which is not the case here}
We note that for {$\Delta{}T \simlt 3.5$ msec} the coherent model is still favoured, as the probability distribution still contains a sufficient probability of intersecting the coherent signal manifold.
Although this plot is merely representative of a single test of coherence and particular details will vary with the signals and datasets, we have specifically used identical signals so that the PDF should peak at the same value in all but the time parameter. This should correspond to a situation where it is extremely difficult to determine whether or not the multiple signals are incoherent or coherent and provide a challenging test of the method.

We consider now a second test of this method. A situation which commonly arises is the presence of a ``glitch'' or instrumental artefact in a single interferometer which is not present in the others in the network. This situation is handled in the case of an incoherent search by checking that corresponding triggers, with consistent physical parameters, exist in all interferometers. As a useful sanity check for the coherence test, we examined the case in which the ``glitch'' has exactly the functional form of a gravitational wave from an in-spiral signal, but is present only in one instrument. In this case, the Bayes factor of the coherent model against the Gaussian noise model is elevated, however our coherence test should allow us to exclude an event such as this due to the lack of a consistent signal in the other detectors.

Table \ref{tab:Honly} displays the detailed results of performing the coherent and incoherent analyses on a signal with an SNR of 9.8 injected only into LIGO-Hanford (H) simulated data. It is notable that even using the coherent model, the signal causes an elevated Bayes factor to be found, as there is some set of parameters which give a compromise signal in the network of detectors. This gives us the undesirable situation where $B_{S,N}$, {the Bayes factor of signal against Gaussian noise, Eq.~(\ref{e:Bsn})},  would be triggered by an event in a single detector. However, by performing the coherence test, we can see that the incoherent model is favoured by a factor {$\approx 10^3$}
to the coherent model, indicating that this situation is unlikely to be a true coherent gravitational wave, and so it can be safely ruled out.

This also works in the case of a two-detector network, although in this case there is a much stronger possibility that a signal may be observed in only one of the two detectors, and so the corresponding analysis infers that there is only a 2.9 times greater chance of the incoherent model than the coherent one.

\begin{table}
\begin{tabular}{|c|cc|cc|c|}
\hline{}
Instruments 	& \multicolumn{2}{c|}{$\log_e{}Z$} 				& \multicolumn{2}{c|}{$\log_e{}B_\mathrm{S,N}$}			& $\log_e B_\mathrm{coh,inc}$		\\
\hline
			& $\H_\mathrm{inc}$		&	$\H_\mathrm{coh}$	& $\H_\mathrm{inc}$		&	$\H_\mathrm{coh}$	&							\\
\hline
H			&	--				& -5950.45			&	--				&  45.78				&	--						\\
L			&	--				& -6106.47			&	--				&  0.39 				&	--						\\
V			&	--				& -6059.44			&	--				& -0.80 				&	--						\\
\hline
HL			& -12056.93			& -12058.00			& 46.16				& 45.09				& -1.07						\\
HLV			& -18116.37			& -18123.29			&  45.36				& 7.52				& -6.92						 \\
\hline
\end{tabular}
\caption{\label{tab:Honly}Results of performing {the coherence test on a signal injected only into the LIGO-Hanford simulated data set.} The test successfully rules out a signal which does not appear in more than one detector, {despite the coherent signal vs. noise comparison ($B_{S,N}$) still favouring the signal model for HLV and HL observations.}
}
\end{table}

This test of coherence gives us a powerful means of distinguishing coherent from incoherent events, which can be used to quantify the additional confidence that we achieve through the use of a network. This result follows naturally from the precise statement of the hypotheses using the conceptual and computational framework that we have developed and demonstrates the power that Bayesian methodology can bring.

\subsubsection{Parameter resolution}

\begin{figure*}
\begin{tabular}{ccc}
\includegraphics[width=0.32\textwidth]{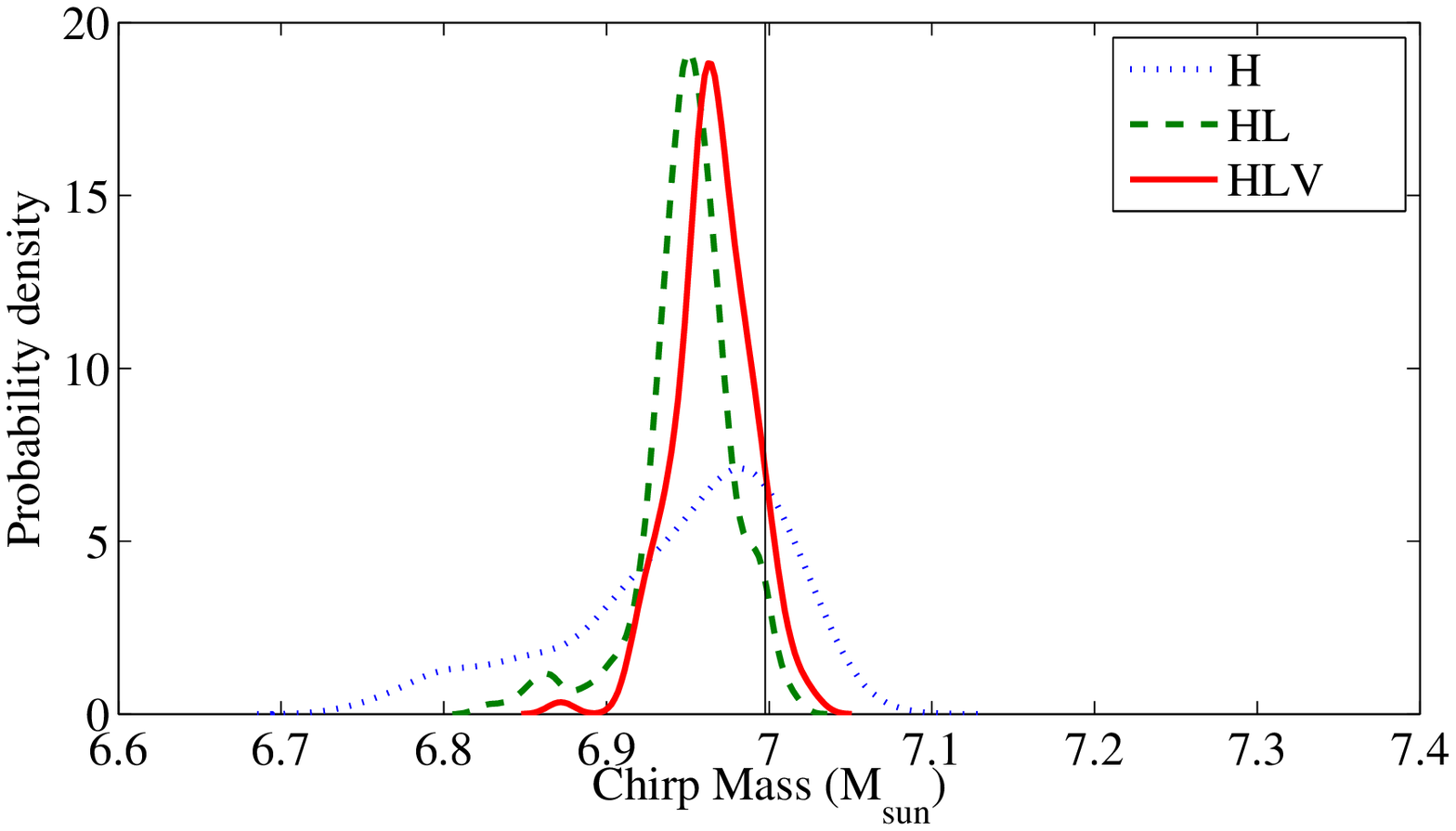} \hspace{-0.25cm} &
\includegraphics[width=0.32\textwidth]{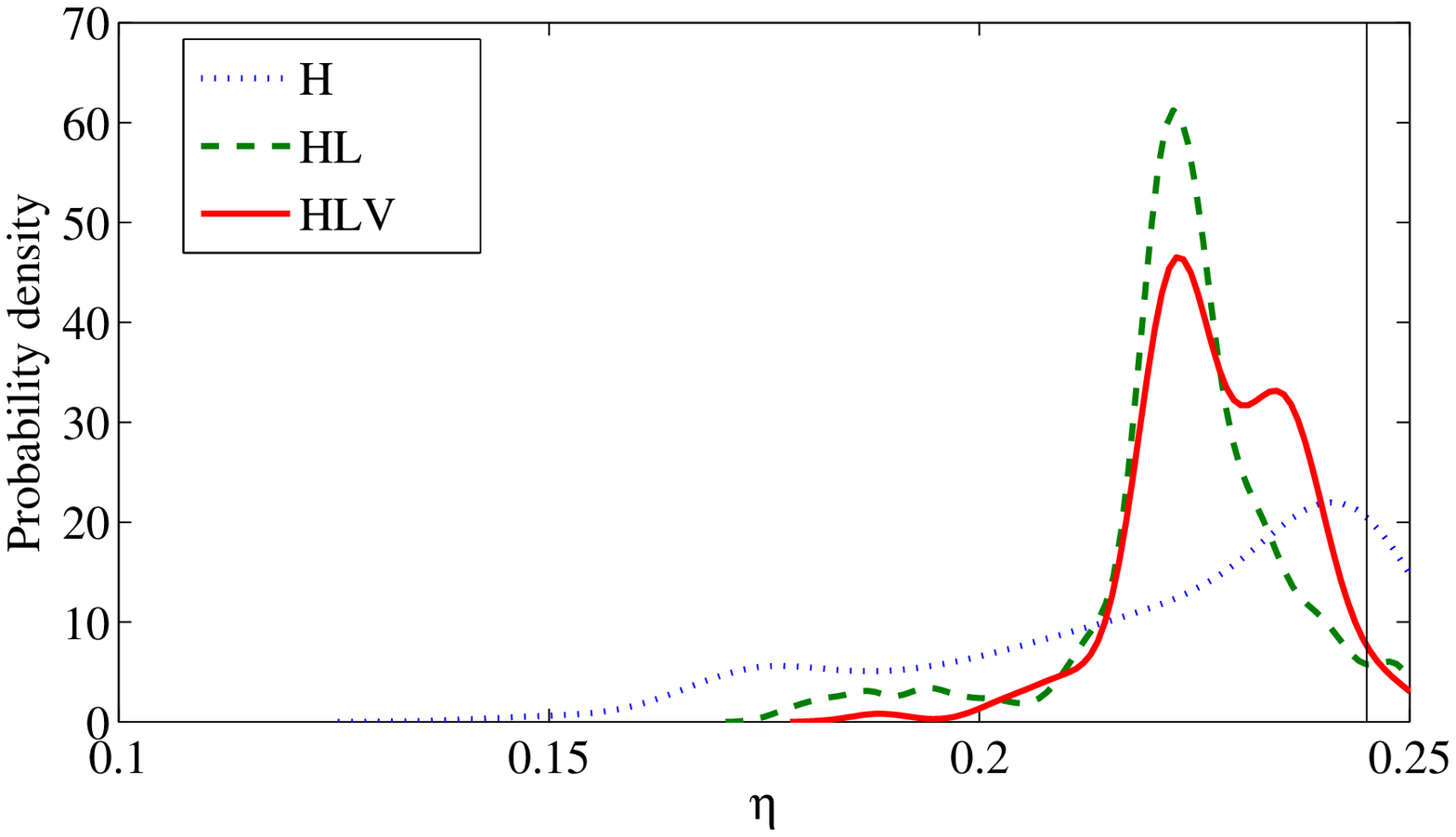} \hspace{-0.25cm} &
\includegraphics[width=0.32\textwidth]{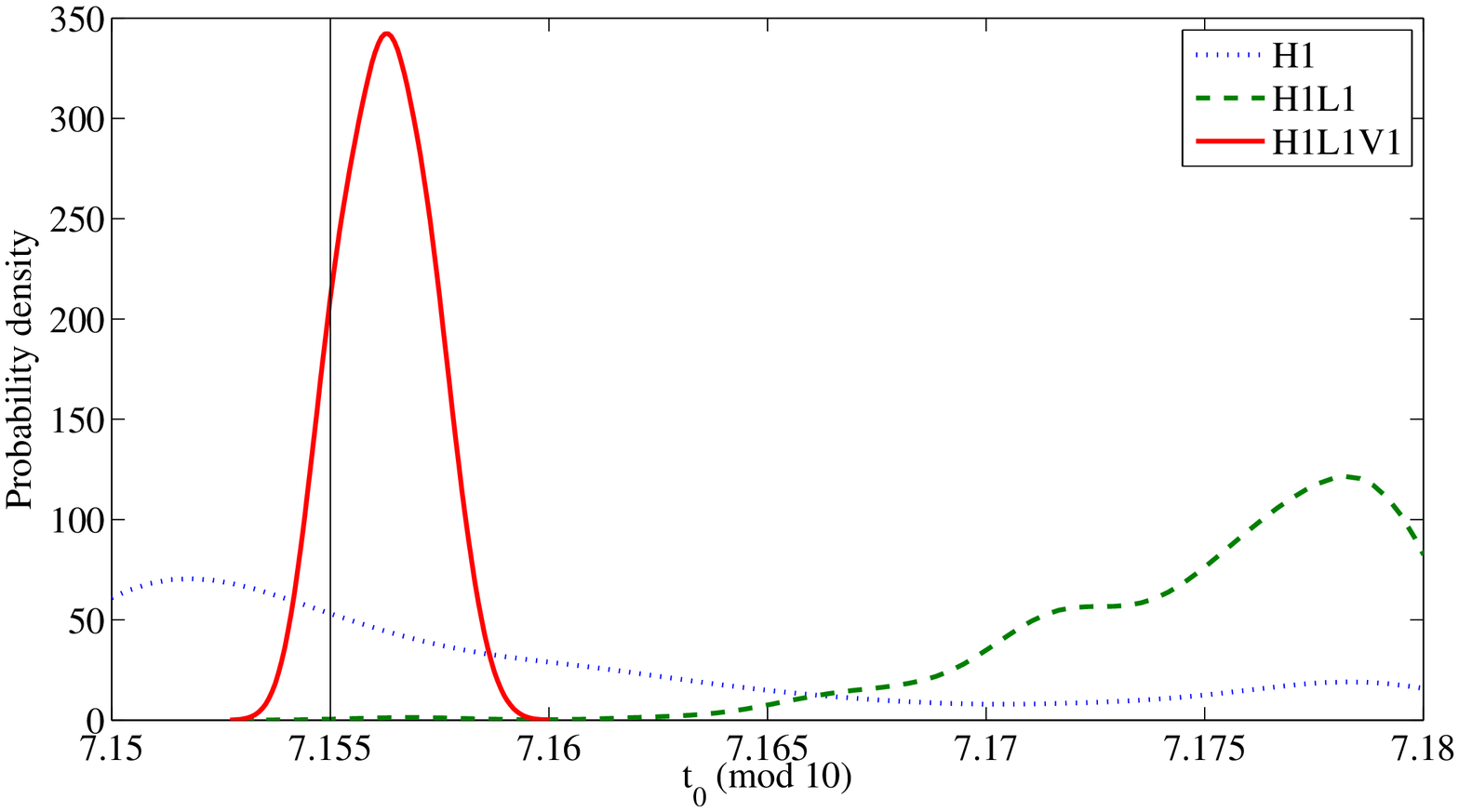}\\
\includegraphics[width=0.32\textwidth]{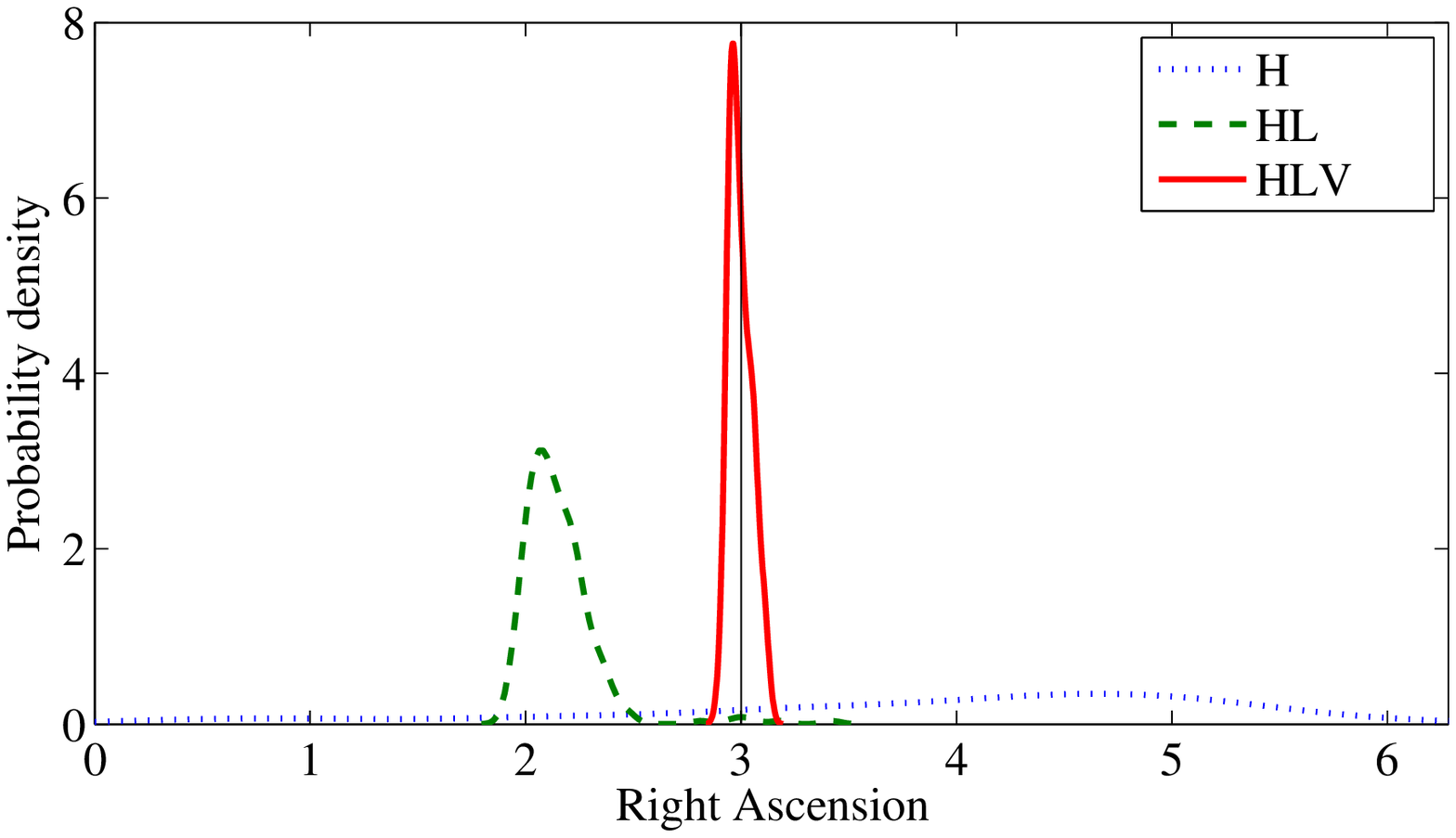} \hspace{-0.25cm} &
\includegraphics[width=0.32\textwidth]{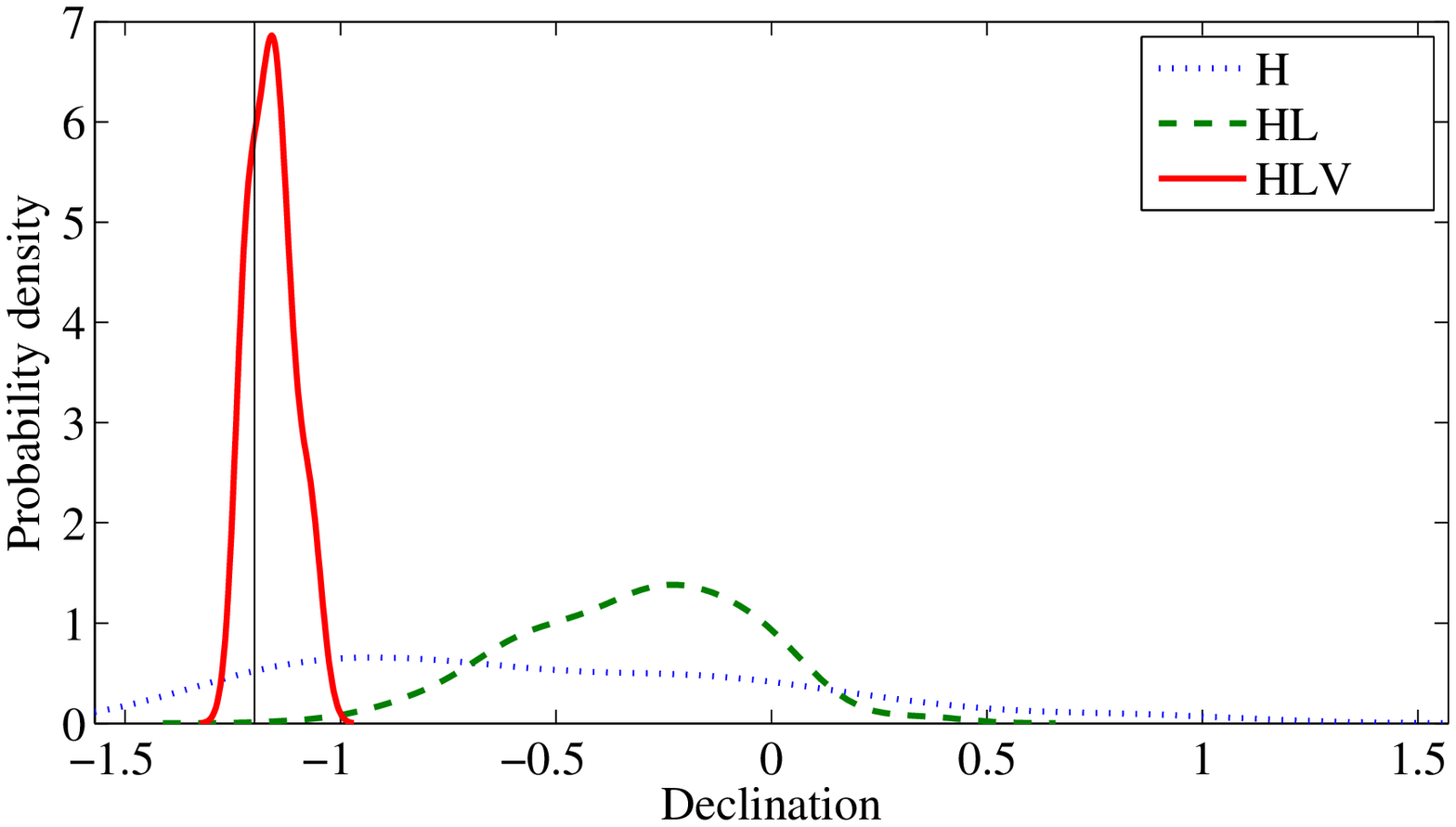} \hspace{-0.25cm} &
\includegraphics[width=0.32\textwidth]{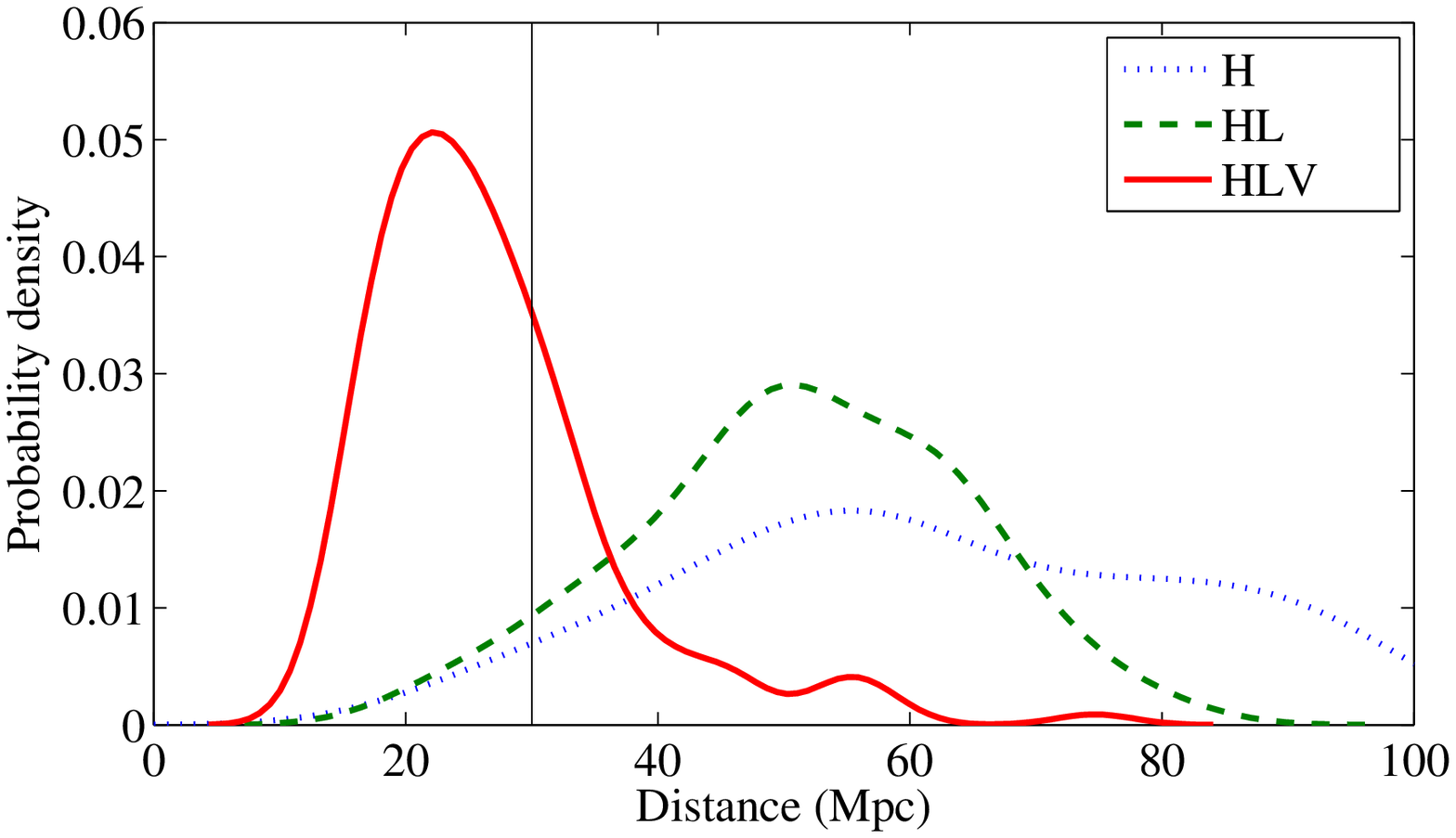} \\
\includegraphics[width=0.32\textwidth]{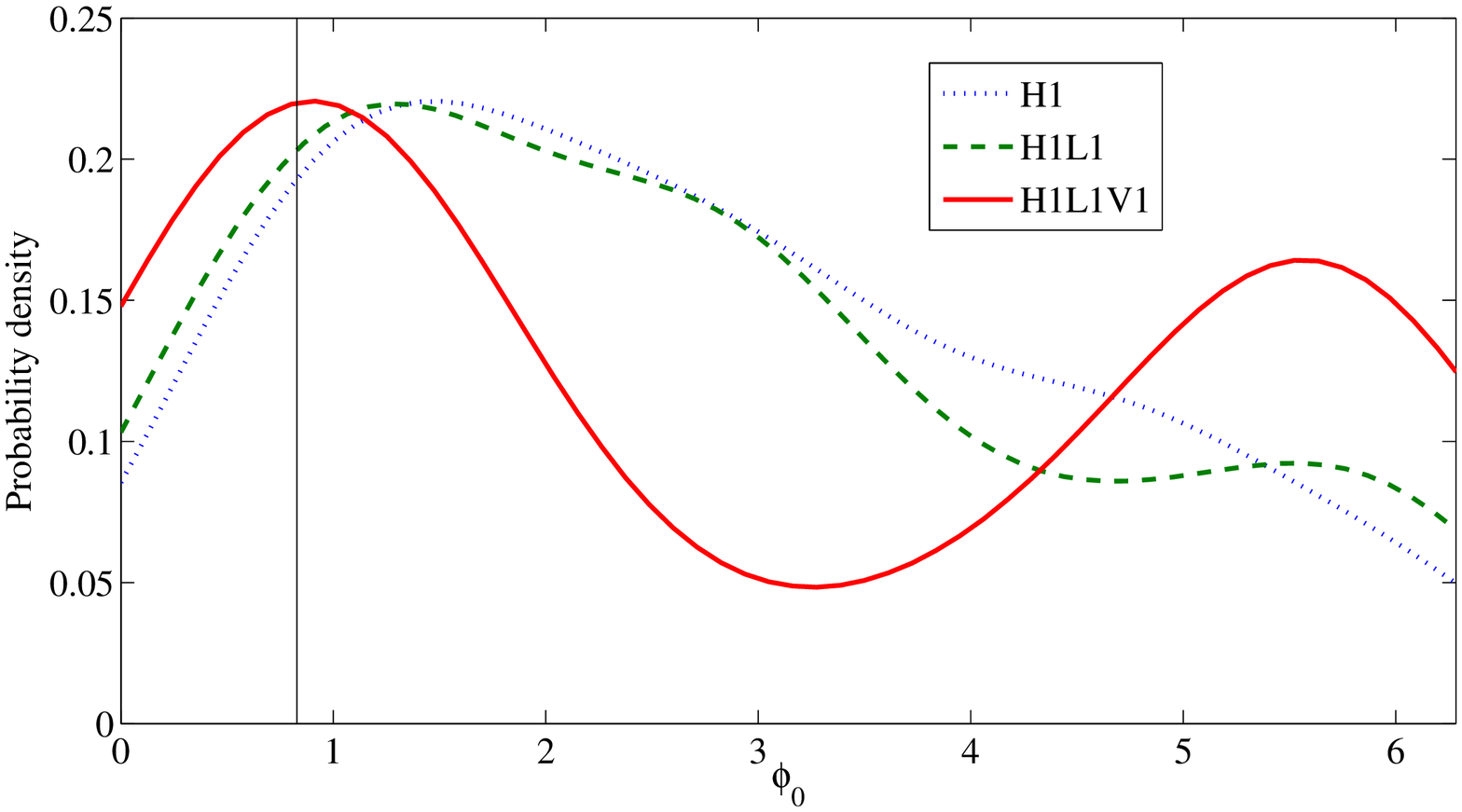} \hspace{-0.25cm} &
\includegraphics[width=0.32\textwidth]{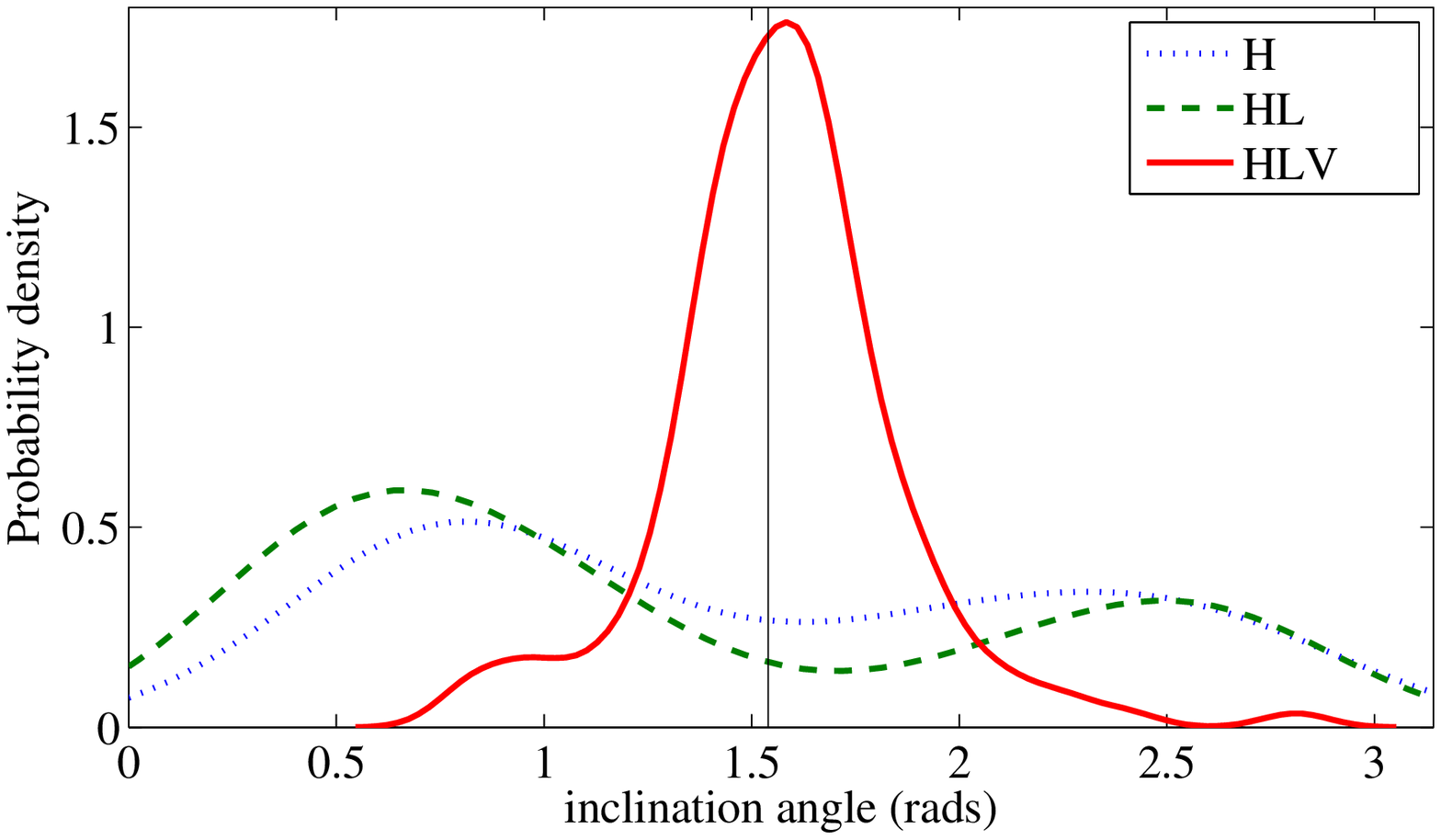} \hspace{-0.25cm} &
\includegraphics[width=0.32\textwidth]{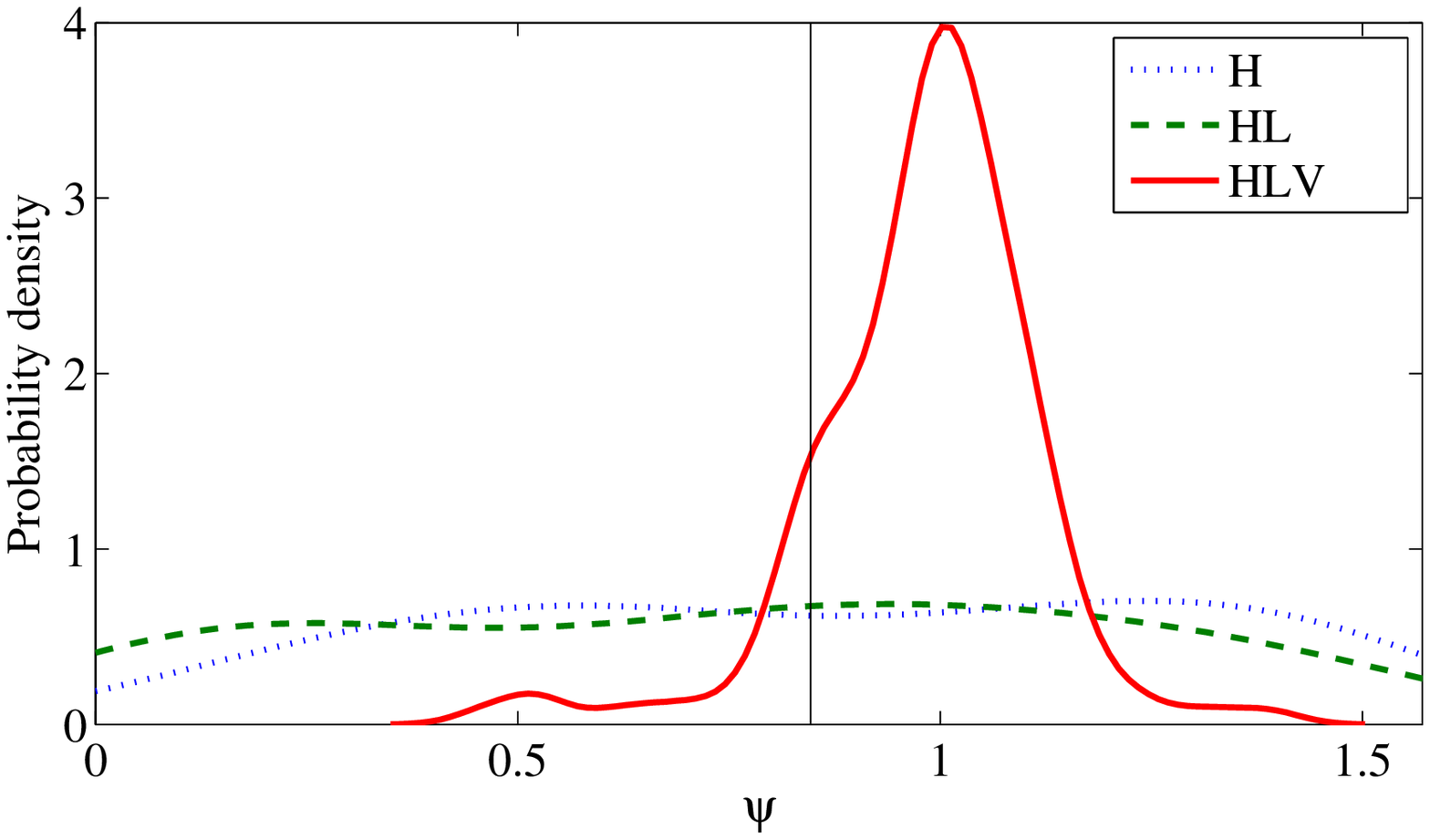} \\

\end{tabular}
\caption{\label{fig:network_params} {The one-dimensional marginalised posterior probability density functions of the nine parameters that describe a gravitational wave in-spiral signal from a circular binary of non-spinning compact objects. The plots show the effect of the coherent network analysis on the estimation of the parameters.}
The true values of each parameter is indicated by a vertical, black line. {The signal was injected in simulated Gaussian and stationary noise representing the LIGO-Hanford, LIGO-Livingston and Virgo instruments, with an optimal signal-to-noise ratio of 9.3 (H), 12.8 (HL) and 14.4 (HLV), respectively.} As more detectors are added to the network, the parameters of the signal become better constrained, with three detectors being necessary to fully resolve all the signal parameters. Lines represent kernel density estimates of the parameters, based on the samples from the PDF generated as in section \ref{ss:Posterior}. Edge effects from the smoothing function are responsible for lowered density estimates near the edges in the distributions of $\phi_0$ and $\eta$. The density estimation was performed using the Matlab function ksdensity.}
\end{figure*}

We have shown in Section~\ref{ss:Posterior} that from the output of the nested sampling algorithm for the evidence/Bayes factor computation one can construct at no additional computational costs the {marginalised} posterior PDFs on the unknown source parameters. Here we show an example of the evaluation of such PDFs with nested sampling as a function of the number of instruments in the network, applied to the coherent observations of an in-spiral binary signal. We consider a system with an optimal signal-to-noise ratio of 9.3, 12.8 and 14.4 respectively in the simulated network configurations of Hanford only, Hanford-Livingston and Hanford-Livingston-Virgo. The actual values of the parameters used  for the injection are shown by the black vertical lines in Figure~\ref{fig:network_params}.

Each instrument measures essentially two independent quantities -- an amplitude and a phase -- as a function of time. As the duration of an in-spiral is negligible in comparison to the period of rotation of the Earth, there is no observable evolution of the antennae response functions during the period of observation (from which one would otherwise reconstruct the source location in the sky). 
{From the signal strain and the time of arrival of the gravitational-wave burst, one} must infer the parameters $\Mc$, $\eta$, $t_{0}$, $D_L$, $\alpha$, $\delta$, $\psi$ and $\iota$. The chirp mass $\Mc$ (and to lesser extent the symmetric mass ratio $\eta$) %contributes to 
{determines} the phase evolution of the signal, which provides a large amount of information to constrain this parameter independent of the others listed here. However, in the remaining parameters a large degree of degeneracy is present, producing correlated joint posterior {PDFs.} 
For the case of observations with one interferometer, this manifests as a broad distribution in the one dimensional marginal PDFs shown in Figure \ref{fig:network_params}. When there is insufficient information available to determine these parameters, the posterior PDFs can be influenced more strongly by the prior distribution, {as it is the case in particular for $D_L$, $\alpha$, $\delta$, $\psi$ and $\iota$.}

With the addition of a second, independent detector at a geographically different location, the possible sky locations and times of arrival are strongly constrained to lie on the surface described in Section \ref{ss:MCMCcustom}, but there is still substantial uncertainty in the marginal distributions for the chosen parameters. This is finally broken when a third detector is added to the network (red line in Figure \ref{fig:network_params}), which allows the sky location to be determined uniquely, along with the remaining parameters.
In this figure we can see the evolution of the posterior PDFs with additional observations, as described above.
For this injection, an inclination was used which placed the orbital plane of the binary almost edge-on to the line of sight to the Earth, meaning that only one polarisation was detectable. This is reflected in the under-estimation of the distance to the system for a network of less than 3 detectors, as there are many positions which are not edge-on which lead to a higher overall observed signal amplitude, and therefore must be located farther away.

This provides an example of the necessity of multiple detectors, operating as a network, if we are to make full use of the astrophysical information carried {by} gravitational waves. By underdetermining the parameters of the signal, biases or at least additional uncertainties may be introduced into our conclusions about the nature of observed sources.

\section{Conclusions}

By taking a Bayesian approach to the analysis of data for the detection and characterisation of in-spiral signals, we have been able to implement a conceptually simple yet flexible framework for drawing inference from observations. Although the Bayesian formalism calls for the evaluation and integration of high-dimensional likelihood functions, we have shown that the nested sampling 
{technique} provides us with a means to both search for and estimate the parameters of a signal. Further work remains to be done in improving the efficiency and reliability of detection at low masses, but the particular implementation that we describe here provides a solid basis for these future improvements.

We have used our implementation to demonstrate the power of Bayesian model selection in classifying putative gravitational wave signals, through the use of the coherence test described in Section \ref{sec:coherent}. The coherence test goes some way to implementing a robust and Bayesian defence against glitches present in gravitational wave data which do not resemble coherent gravitational waves by providing an internal consistency check within the detector network. Tests of this kind may provide a useful new additional discriminator when analysing candidate gravitational waves, and are only achievable through the treatment of the detector network in a coherent way. Notably, this test is only possible through the use of the Occam factor and the comparison of probability distributions which differ in their dimensionality, and could not be possible with point estimates of maximum likelihood. Indeed, the maximum likelihood of the coherent and incoherent models can be trivially shown to be identical. In addition to such tests, it would also be desirable to model the detector data in a way which is non-stationary or Gaussian, but this is beyond the scope of this work. Furthermore, the use of the Bayesian parameter estimation framework is invaluable in inferring the signal parameters, and produces the full posterior probability distribution, not just maximum likelihood points and estimates of variance. The covariance and interdependence of the parameters does not prevent us from calculating consistent joint PDFs on the full parameter space even when point estimates have little meaning, and this allows us to see the benefit of using a network of detectors in a coherent fashion.

{One of} the main benefits in the use of the nested sampling and Bayesian evidence approach is the ease with which it can be extended or adapted to different signal models with minimal changes needed to the implementation. In other work past and ongoing, we have shown how this method may be applied to the discrimination between candidate waveforms when comparing to a numerical relativity simulation; how one may place bounds on the Compton wavelength of the graviton given a single or multiple observations of an in-spiral signal, and testing the effects of including spins in the detection and estimation of in-spiral signals \cite{AylottEtAl:2009,WDP:lambdaG}\,.

Future work on the implementation of the core algorithm will focus on achieving a better reliability and efficiency in the sampling of the parameter space, which should lead to further improvements in the performance in a real world situation, and on testing on the full set of waveform approximants used in present searches for coalescing binary systems. The work presented here forms the basis of the \texttt{ lalapps\_inspnest } program, which is can be found in the LALApps software distribution, released under the terms of the GNU General Public Licence \cite{lalapps}.

\acknowledgments

This work has benefited from many discussions with members of the LIGO Scientific Collaboration, {in particular we would like to thank Martin Hendry, Ilya Mandel and Christian R\"over for their comments and suggestions}.
The simulations presented in this paper were carried out on the Tsunami Beowulf clusters of the University of Birmingham and the Atlas cluster of the AEI Hannover. This work has been supported by the UK Science and Technology Facilities Council. This paper has been assigned the LIGO Document Number LIGO-P0900117

\appendix
\section{Definitions and conventions}\label{a:definitions}

We provide the definitions and conventions that we have adopted to link the continuous representation of a time series which is used in the main body of the paper with the discrete representation. The latter is what is actually used for applications, and corresponds to the expressions implemented in the \texttt{ lalapps\_inspnest }  software. This appendix provides also a mapping of the notation used in our previous papers~\cite{VeitchVecchio:2008a, VeitchVecchio:2008b} and the one used here.

Consider a time series $a(t)$ sampled at time intervals $\Delta t = 1/(2 f_\mathrm{Ny})$, where $f_\mathrm{Ny}$ is the \emph{Nyquist frequency}, for a total observation time $T$. The total number of data samples is therefore $N_p = {T}/{\Delta t} = 2 T f_\mathrm{Ny}$. Given a generic real function $a(t)$, our conventions for the Fourier Transform are~\footnote{We are adopting here the convention used by the GW data analysis work and the LIGO Scientific Collaboration, which is also the engineering convention}:
\be
\tilde a(f) = \int_{-\infty}^{+\infty} a(t)e^{-2\pi i f t} dt\,,
\label{e:cFT}
\ee
and the inverse transform is
\be
a(t) = \int_{-\infty}^{+\infty} \tilde a(f)e^{+2\pi i f t} df\,.
\label{e:cinvFT}
\ee
The data points at time $t_j$ and frequency $f_k$ are therefore:
\ba
a(t_j) & = & a(j\Delta t) = a_j
\\
\tilde a(f_k) & = & \tilde a(k/T) = \Delta t\times \tilde a_k\,,
\label{e:afk}
\ea
where we have defined the Fourier series as:
\ba
\tilde a_k & = & \sum_j a_j\, e^{-2\pi i jk/N_p}\,,
\label{e:ak}
\\
 a_j & = & \frac{1}{N_p}\sum_j \tilde a_k\, e^{+2\pi i jk/N_p}\,.
\label{e:aj}
\ea
In the following we will indicate with $\tilde a(f_k)$ the (dimension-full) approximation to the Fourier Transform of $\tilde a(f)$ at frequency $f = f_k$.

We consider now the statistical properties of the noise. The \emph{one-sided} noise spectral density $S(f)$ is defined in the continuous case as:
\be
S(f) = 2\int_{-\infty}^{+\infty} \m n(t + \tau) n(\tau) \M e^{-i 2\pi f t} dt\,,
\label{e:Sfdef}
\ee
which yields
\be
\m \tilde n(f) \tilde n^*(f')\M = \frac{1}{2} S(f) \delta (f-f') = S^{(2)}(f) \delta (f-f')\,,
\label{e:nvar}
\ee
where the factor $1/2$ comes from the fact that $S(f)$ is the one-sided noise spectral density; this is related to the \emph{two-sided} noise spectral density $S^{(2)}(f)$ by
\be
S^{(2)}(f) =  \frac{1}{2} S(f)\,.
\ee

If we explicitly write the real and imaginary part of the noise contribution in the Fourier domain, which is the notation adopted in~\cite{VeitchVecchio:2008a, VeitchVecchio:2008b}  as
\be
\tilde n(f_k) = \tilde x(f_k) + i \tilde y(f_k)\,,
\ee
and we substitute into Eq.~(\ref{e:nvar}), we obtain:
\be
\m |\tilde n(f_k)|^2 \M = \m |\tilde x(f_k)|^2 \M + \m |\tilde y(f_k)|^2 \M = \frac{T}{2} S(f_k)\,,
\ee
where the first equality comes from the fact that $\tilde x(f_k)$ and $\tilde y(f_k)$ are independent. In terms of the elements of the Fourier series one has:
\ba
\Delta t^2\m |\tilde n_k|^2\M & = & 
\Delta t^2 \left[\m |\tilde x_k|^2\M + \m |\tilde y_k|^2\M \right]\,.
\nonumber\\
& = & \frac{T}{2} S(f_k)
\label{e:nvar1}
\ea
If one defines  
\ba
\sigma_k^2 & = & \m |\tilde n_k|^2\M
\nonumber\\
\zeta_{k}^2 & = & \m |\tilde x_k|^2\M = \m |\tilde y_k|^2\M
\nonumber\\
\sigma_k^2 & = & 2 \zeta_{k}^2
\ea
the variance of the (complex) noise and the real and imaginary part of the Fourier series elements, then
\ba
S(f_k) & = & 
2\frac{\Delta t^2}{T} \sigma_k^2 = 2\frac{\Delta t}{N} \sigma_k^2\,,
\nonumber\\
& = & 4 \frac{\Delta t}{N} \zeta_k^2\,.
\label{e:Sfk}
\ea

Let us now considered the usual inner product $(.|.)$~\cite{CutlerFlanagan:1994} between two (real) functions $a$ and $b$ and its approximation in the finite case:
\ba
(a|b) & = & 2\int_0^{\infty}\frac{\tilde a(f)\tilde b^*(f) + \tilde a^*(f)\tilde b(f)}{S(f)}\,df
\\
& \approx & \frac{2}{T}\sum_{k>0} \frac{\tilde a(f_k)\tilde b^*(f_k) + \tilde a^*(f_k)\tilde b(f_k)}{S(f_k)}
\\
& \approx & \sum_{k>0} \frac{\tilde a_k\tilde b^*_k + \tilde a^*_k\tilde b_k}{\sigma_k^2}
\label{e:inner}
\ea
The optimal signal-to-noise ratio is just the square root of the norm of $h$, and using Eq.~(\ref{e:inner}) it yields:
\ba
(h|h) & = & 4\int_0^{\infty}\frac{|\tilde h(f)|^2}{S(f)}\,df\,,
\nonumber\\
& \approx & \frac{4}{T}\sum_{k>0} \frac{|\tilde h(f_k)|^2}{S(f_k)}\,,
\nonumber\\
& \approx & 2\sum_{k>0} \frac{|\tilde a_k|^2}{\sigma_k^2}\,.
\ea
The likelihood function for the data set $d$ given the model $h$, Eq.~(\ref{e:L}) therefore becomes
\ba
p(d|h,\H_S) & \propto & \exp\left[-\frac{1}{2} \left(d - h | d-h\right)\right]
\\
& \propto & \exp\left[-\frac{2}{T} \sum_{k>0} \frac{|\tilde d(f_k) - \tilde h(f_k)|^2}{S(f_k)}\,\right]\,,
\ea
and this is the expression used in the software implementation through Eqs.~(\ref{e:afk}), (\ref{e:ak}) and~(\ref{e:Sfk}).

\end{document}